\documentclass[pre,twocolumn,showpacs,showkeys,superscriptaddress,preprintnumbers,floatfix]{revtex4-1}

\usepackage{etex}
\usepackage{ifpdf}
\usepackage{hyperref}
\usepackage{dcolumn}
\usepackage{url}
\usepackage{amsmath}
\usepackage{amscd}
\usepackage{amsfonts}
\usepackage{amssymb}
\usepackage{bm}   
\usepackage{bbm}
\usepackage{verbatim}
\usepackage{stmaryrd}
\usepackage{amsthm}
\usepackage{xcolor}
\usepackage{setspace}
\usepackage{braket}

\usepackage{tikz}
\usetikzlibrary{arrows}
\usetikzlibrary{automata}
\usetikzlibrary{decorations.pathreplacing}
\usetikzlibrary{positioning}
\usetikzlibrary{plotmarks}
\usetikzlibrary{calc}
\usetikzlibrary{patterns}
\usetikzlibrary{external}
\usepackage{pgfplots}
\pgfplotsset{compat=newest}

\usepackage{graphicx}

\theoremstyle{plain}    
\theoremstyle{plain}    
\theoremstyle{plain}    
\theoremstyle{plain}    
\theoremstyle{plain}    
\theoremstyle{plain}    
\theoremstyle{plain}    
\theoremstyle{plain}    
\theoremstyle{plain}    
\theoremstyle{plain}    
\theoremstyle{plain}    
\theoremstyle{plain}

\newcommand{\CausalState}   { \mathcal{S} }

\newcommand{\forward}{+}
\newcommand{\reverse}{-}
\newcommand{\forwardreverse}{\pm} 

\newcommand{\FutureCausalState} { {\CausalState}^{\forward} }

\newcommand{\PastCausalState}   { {\CausalState}^{\reverse} }

\newcommand{\lastindex}[2]{
  \edef\tempa{0}
  \edef\tempb{#2}
  \ifx\tempa\tempb
    \edef\tempc{#1}
  \else
    \edef\tempa{0}
    \edef\tempb{#1}
    \ifx\tempa\tempb
      \edef\tempc{#2}
    \else
      \edef\tempc{#1+#2}
    \fi
  \fi
  \tempc
}

\newcommand{\CSjoint}[1][,]{
   \edef\tempa{:}
   \edef\tempb{#1}
   \ifx\tempa\tempb
      \ensuremath{\FutureCausalState\!#1\PastCausalState}
   \else
      \ensuremath{\FutureCausalState#1\PastCausalState}
   \fi
}

\newif\ifpm
\edef\tempa{\forwardreverse}
\edef\tempb{\pm}
\ifx\tempa\tempb
   \pmfalse
\else
   \pmtrue
\fi

\renewcommand{\H}{\operatorname{H}}

\newcommand{\kB}{k_\text{B}}  

\begin{document}

\title{Shortcuts to Erasure:\\
Fast, Efficient, and Accurate\\
Thermodynamic Computing}
\title{Shortcuts to Thermodynamic Computing:\\
The Cost of Fast and Faithful Erasure}

\author{Alexander B. Boyd}
\email{abboyd@ucdavis.edu}
\affiliation{Complexity Sciences Center and Physics Department,
University of California at Davis, One Shields Avenue, Davis, CA 95616}

\author{Ayoti Patra}
\email{ayotipatra@gmail.com}
\affiliation{Department of Physics, University of Maryland, College Park,
Maryland 20742, USA}

\author{Christopher Jarzynski}
\email{cjarzyns@umd.edu}
\affiliation{Department of Physics, University of Maryland, College Park,
Maryland 20742, USA}
\affiliation{Department of Chemistry and Biochemistry, University of Maryland, College Park,
Maryland 20742, USA}
\affiliation{Institute for Physical Science and Technology, University of
Maryland, College Park, Maryland 20742, USA}

\author{James P. Crutchfield}
\email{chaos@ucdavis.edu}
\affiliation{Complexity Sciences Center and Physics Department,
University of California at Davis, One Shields Avenue, Davis, CA 95616}

\date{\today}
\bibliographystyle{unsrt}

\begin{abstract}
Landauer's Principle states that the energy cost of information processing must
exceed the product of the temperature and the change in Shannon entropy of the
information-bearing degrees of freedom. However, this lower bound is achievable
only for quasistatic, near-equilibrium computations---that is, only over
infinite time. In practice, information processing takes place in finite time,
resulting in dissipation and potentially unreliable logical outcomes. For
overdamped Langevin dynamics, we show that counterdiabatic potentials can be
crafted to guide systems rapidly and accurately along desired computational
paths, providing shortcuts that allows for the precise design of finite-time
computations. Such shortcuts require additional work, beyond Landauer's bound,
that is irretrievably dissipated into the environment. We show that this
dissipated work is proportional to the computation rate as well as the square
of the information-storing system's length scale. As a paradigmatic example, we
design shortcuts to erase a bit of information metastably stored in a
double-well potential. Though dissipated work generally increases with erasure
fidelity, we show that it is possible perform perfect erasure in finite time
with finite work. We also show that the robustness of information storage
affects the energetic cost of erasure---specifically, the dissipated work
scales as the information lifetime of the bistable system. Our analysis exposes
a rich and nuanced relationship between work, speed, size of the
information-bearing degrees of freedom, storage robustness, and the difference
between initial and final informational statistics.
\end{abstract}


\preprint{arxiv.org:1812.XXXXX [cond-mat.stat-mech]}

\maketitle



\setstretch{1.1}
\section{Introduction}
\label{sec:Intoduction}

Information processing requires work. For example, no less than $\kB T \ln2$ of
work must be supplied in order to erase a single bit of information at
temperature $T$ \cite{Land61a}. More generally, Landauer's Principle bounds the
work investment by the change in the memory's Shannon entropy \cite{Parr15a}:
\begin{align}
\langle W \rangle \geq \kB T \ln 2 \left( \H[Y_0] - \H[Y_\tau] \right)
  ~.
\label{eq:LandauerBound}
\end{align} 
Here, $Y_0$ and $Y_\tau$ are random variables describing initial and final
memory states, and $\H[Y]=-\sum_{y} \Pr(Y=y) \log_2 \Pr(Y=y)$ denotes the
uncertainty in bits of a random variable $Y$.

Mathematically, information processing is described by a communication channel
\cite{Shan48a} that maps an initial distribution $\Pr(Y_0)$ to a final
distribution $\Pr(Y_\tau)$. Physically, a memory is realized by a system whose
thermodynamically-metastable states encode logical states $\{y\}$. The simplest
example is a Brownian particle in a double-well potential, with two deep wells
representing the $y=0$ and $y=1$ states of a single bit of information. More
generally, the collection of all possible memory states $\mathcal{Y}=\{y\}$
represents a mesoscopic coarse-graining of the space of explicit physical
microstates $\mathcal{X}=\{x\}$ of the memory device. Information processing is
implemented by varying the system's energy landscape so as to drive the flow of
probability between memory states in a controlled fashion, to achieve a desired
computation.

A computation can be implemented to achieve the Landauer bound,
Eq.~(\ref{eq:LandauerBound}), by varying the energy landscape infinitely
slowly, so that the system remains in metastable equilibrium from beginning to
end \cite{Boyd17a,Garn15}. Such quasistatic computations, however, take
infinitely long to implement. For computations performed in finite time the
underlying physical system is driven out of equilibrium, resulting in the
irretrievable dissipation of energy into thermal surroundings. The problem of
minimizing this dissipation has recently been explored in the linear response
regime, using the tools of geometric thermodynamic control \cite{Zulk12a,
Zulk14a, Zulk15a, Ging16a}.

In the present work, we consider the separate problem of how to implement a
computation rapidly and reliably. That is, we study how to design protocols for
varying the energy landscape of a system, so as to produce a desired
computation in a given time interval, no matter how short its duration $\tau$.
In effect, we place a premium on speed of computation rather than energy
efficiency, although we then proceed to analyze the energetic costs of rapid
computation.

To achieve rapid and precisely-controlled information processing, we use
recently developed tools from the field of \emph{shortcuts to adiabaticity}
\cite{Torrontegui13}. Specifically, we apply generic methods of counterdiabatic
control of classical overdamped systems \cite{Patr17a}, which were inspired by
pioneering experiments on the engineered swift equilibration of a Brownian
particle \cite{Mart16a}. The results we obtain are not limited to the linear
response regime---they remain valid even when the system is driven far from
equilibrium during the information processing.

For concreteness, we show how to apply counterdiabatic control to erase a
single bit of information rapidly and accurately, but our approach generalizes
to other cases of information processing. To embed the memory states
$\mathcal{Y}$ physically, we consider a one-dimensional position space
$\mathcal{X}$ governed by overdamped Fokker-Planck dynamics. The energy
landscape at the beginning and end of the protocol is the double-well potential
shown in Fig. \ref{fig:MetastableDistributions}, with a barrier sufficiently
high to prevent the leakage of probability between the two wells. Thus, the
landscape provides a means of storing information in metastable mesoscopic
states.  As we will show, counterdiabatic control of the potential can be used
to drive any initial distribution over the memory states to any desired final
distribution in finite time---in fact, arbitrarily rapidly.  Mirroring results
in geometric control, we show that the work required to perform this
counterdiabatic process decomposes into a change in free energy, which captures
Landauer's change in state space, plus an additional contribution that scales
as the rate of computation and the square of the length scale of the
information-storing potential \cite{Zulk14a}. This additional work is
proportional to the global entropy production and so quantifies thermodynamic
inefficiency.

Our approach reveals additional trade-offs beyond that between computation
rate, length scale, and thermodynamic efficiency. We show that dissipation also
increases with the difference between initial and final bit distributions of
the computation and with the robustness of information storage. In this way, we
give a more complete picture of information processing beyond Landauer's bound.
Rather than a tradeoff between information and energy, more complex tradeoffs
are revealed between information, energy, statistical bit-bias difference,
speed, size of the memory states, and information robustness.

\section{Thermodynamic Computing}
\label{sec:Physical Computations}

What is physical computing? At the outset, information must be encoded in
collections of microscopic states $\mathcal{X}$ of a physical system. Let
$\mathcal{Y}$ denote these information-containing microstate groups---the
accessible \emph{memory states} \cite{Benn03,Deff2013}. By manipulating the
physical system, a microstate collection evolves, transforming the information
it contains. Generally, an information processor has only partial control over
and knowledge of the underlying microstates of its physical implementation.  We
now consider how such information processing can be modeled by stochastic
dynamics governed by a controlled potential.

\subsection{Memory States and Symbolic Dynamics}
\label{sec:Memory States  and Symbolic Dynamics}

There are many ways to form memory states out of physical microstates. Here, we
choose a framework for information erasure and general information processing
in which the physical degrees of freedom $\mathcal{X}$ participate in
metastable equilibria. Each metastable equilibrium is a microstate distribution
that corresponds to a memory state $y \in \mathcal{Y}$. For example, we can
have memory states $\mathcal{Y}=\{0,1\}$, such that they are stable for
intermediate, if not asymptotically long, time scales. The coarse-graining $c:
\mathcal{X} \rightarrow \mathcal{Y}$ of physical states to form the
informational states specifies the memory alphabet $\mathcal{Y}= \{c(x) |x \in
\mathcal{X}\}$. This translates a distribution $\Pr(X_t)$ over physical
microstates $x \in \mathcal{X}$ to a distribution $\Pr(Y_t)$ over informational
states $y \in \mathcal{Y}$. In this way, controlling a physical system
determines not only its raw physical dynamics, but also the \emph{symbolic
dynamics} of the informational states \footnote{Symbolic dynamics is a
long-lived subfield of dynamics systems \cite{Lind95a}. Our use of it here is
relatively simple, highlighting (i) how mesoscopic symbols capture (or not)
collections of microscopic states and (ii) the concern of proper
coarse-graining to locate information storage and processing. Fully deploying
the symbolic dynamics for thermodynamic computing must wait for a different
venue. Early results, however, do develop the symbolic dynamics of
thermally-activated (noisy) systems \cite{Crut83a}}.

We use random variable notation, $\Pr(X_t)= \{ (x,\Pr(X_t=x)), x \in \mathcal{X}\}$, common in symbolic dynamics \cite{Lind95a}, rather than $\rho(x,t)$, which is more standard in stochastic thermodynamics, due to its specificity and flexibility. The probability of being in microstate $x$ at time $t$ is expressible in both notations $\Pr(X_t=x)=\rho(x,t)$, but the random variable notation works with many different distributions over the same microstate space $\mathcal{X}$. And so, rather than specify many different probability functions, we specify their random variables. Other advantages of this choice is that it readily expresses joint probabilities, such as
$\Pr(X_t=x,X_{t+\tau}=x')$, and entropies:
\begin{align}
\H[X_t]=-\sum_{x \in \mathcal{X}} \Pr(X_t=x) \log_2 \Pr(X_t=x)
  ~.
\end{align}  
While not all of the potential functionality is used in the following, a number
of benefits will follow in due course.

\subsection{Overdamped Fokker-Planck Dynamics}
\label{sec:Overdamped Fokker-Planck Dynamics}

The first challenge of thermodynamic computing is to control a system's
Hamiltonian over the physical degrees of freedom $\mathcal{X}$ such that the
induced microstate distribution $\Pr(X_t)$ at time $t$ matches a desired
distribution $\Pr(X^d_t)$, where $X_t$ and $X^d_t$ are the random variables for
the actual physical distribution and desired physical distribution,
respectively, at time $t$, each realizing states $x \in \mathcal{X}$. The
second challenge, which we come to later, is to associate the microstate
distributions with mesostate distributions that support the desired
information-storing and -processing.

We consider a Hamiltonian controlled via a potential energy landscape $V(x,t)$
over the time interval $t \in(0,\tau)$, where $x\in \mathcal{X}$. We will
demonstrate that one can exactly guide an overdamped Fokker-Planck dynamics in
one dimension along the desired time sequence of distributions $\Pr(X^d_t=x)$,
resulting in a powerful tool for thermodynamic control and information
processing.

In fact, overdamped stochastic systems are a promising and now common framework
for elementary thermodynamic information processing \cite{Gavr16a, Proe16a}.
With a single physical degree of freedom $\mathcal{X}={\Bbb R}$, one information
processing task is to change the initial distribution to a final distribution
in finite time. The actual microstate distribution $\Pr(X_t)$ obeys the Fokker-Planck equation:
\begin{align}
\frac{\partial \Pr(X_t=x)}{\partial t} = \mu \frac{\partial}{\partial x}
  & \left( \Pr(X_t=x)\frac{\partial V(x,t)}{\partial x}\right) \nonumber \\
  & +\mu \kB T \frac{\partial^2 \Pr(X_t=x)}{\partial x^2}
  ~,
\label{eq:FokkerPlanck}
\end{align}
where $V(x,t)$ is the potential energy landscape at time $t$, $T$ is the
temperature of the thermal environment, and $\mu$ is the inverse friction
coefficient. Recall that the Boltzmann equilibrium distribution:
\begin{align}
\Pr(X^\text{eq}_t=x)=\frac{e^{-V(x,t)/\kB T}}{Z(t)}
  ~,
\label{eq:BoltzmannDistFromPotential}
\end{align}
is a stationary distribution for the Fokker-Planck equation if the potential
is held fixed at time $t$. That is, substituting into the righthand side
of Eq. (\ref{eq:FokkerPlanck}) yields:
\begin{align*}
\frac{ \partial \Pr(X^\text{eq}_t=x)}{\partial t}=0
  ~.
\end{align*}

\section{Work Production During Counterdiabatic Protocols}

Next, we identify how the evolution of the physical distribution yields useful
changes in memory states that robustly store a computation's result. We break
the development into two parts. This section considers counterdiabatic
Hamiltonian control of the physical states $x \in \mathcal{X}$ such that they
follow specified distributions $\Pr(X^d_t)$ over the time interval $t \in (0,
\tau)$ \cite{Patr17a}. For the resulting finite-time protocol, we determine the
work production and show that it increases with both the size of the memory
states and the speed of operation, if the overall computational task is fixed.
This holds for any counterdiabatically-controlled computation. The subsequent
section then addresses the particular computational task of information erasure
in a bistable potential well. While the analytical and numerical results there
do not explicitly generalize to other computational tasks, they introduce
general relationships between dissipated work, information storage robustness,
and computation fidelity that hold broadly.

\subsection{Inverse Problem for Thermodynamic Control}
\label{sec:The Inverse Problem in the Thermodynamics of Control}

For a specified potential $V(x,t)$, the Fokker-Planck equation
Eq.~(\ref{eq:FokkerPlanck}) evolves an initial distribution $\Pr(X_0)$ to a
density $\Pr(X_t)$ at any later time in the control interval $t \in (0,\tau)$.
Together, the probability density and potential determine the average energy
expended as work $\langle W \rangle$ by the protocol on the physical
system~\cite{Deff2013}:
\begin{align}
\langle W \rangle = \int_{0}^\tau dt \int_{-\infty}^{\infty}dx \Pr(X_t=x) \partial_t V(x,t)
  ~.
\label{eq:Wdef}
\end{align}

What if, rather than starting with an initial distribution and control
protocol, we are given a desired trajectory of probability distributions
$\Pr(X^d_t)$ over some time interval $t \in [0, \tau]$, a \emph{distribution
trajectory}, and are tasked to determine the control protocol that yields the
trajectory? This challenge---the \emph{inverse problem} of reconstructing
dynamical equations of motion from distributions over trajectories---falls
within purview of state-space reconstruction \cite{Pack80,Crut87a} and
computational mechanics \cite{Crut12a} which provide principled approaches for
inferring generators of observed time series. Broadly speaking, our challenge
here is to reconstruct dynamical equations of motion for evolving distributions
that perform computations and, then, to show how the work cost relates to the
computation's effectiveness. The setting here is both more constrained and more
challenging than state-space reconstruction.

Generally, as with most inverse problems, determining the control protocol from
a distribution trajectory does not lead to a unique solution. Many alternate
dynamics can generate the same observed distributions \cite{Crut08b}. However, Appendix
\ref{app:Uniqueness of Counterdiabatic Protocols} shows, for the specific case
of overdamped Fokker-Planck dynamics in a single dimension $\mathcal{X}=
\mathbb{R}$, that the distribution trajectory $\Pr(X^d_t)$
uniquely determines the control protocol $V(x,t)$ up to a baseline energy
$E(t)$ that is constant in position and so adds no force. Moreover, if
$\Pr(X^d_t)$ characterizes our desired computation then, up to a
readily-recovered change in baseline energy $E(\tau)-E(0)$, the work is
uniquely determined for that computation. Thus, by designing a single protocol
that guides the system along the desired distribution trajectory, we find both
the unique protocol and the unique work investment required for that
trajectory.

\begin{figure*}[tbp]
\centering
\includegraphics[width=2\columnwidth]{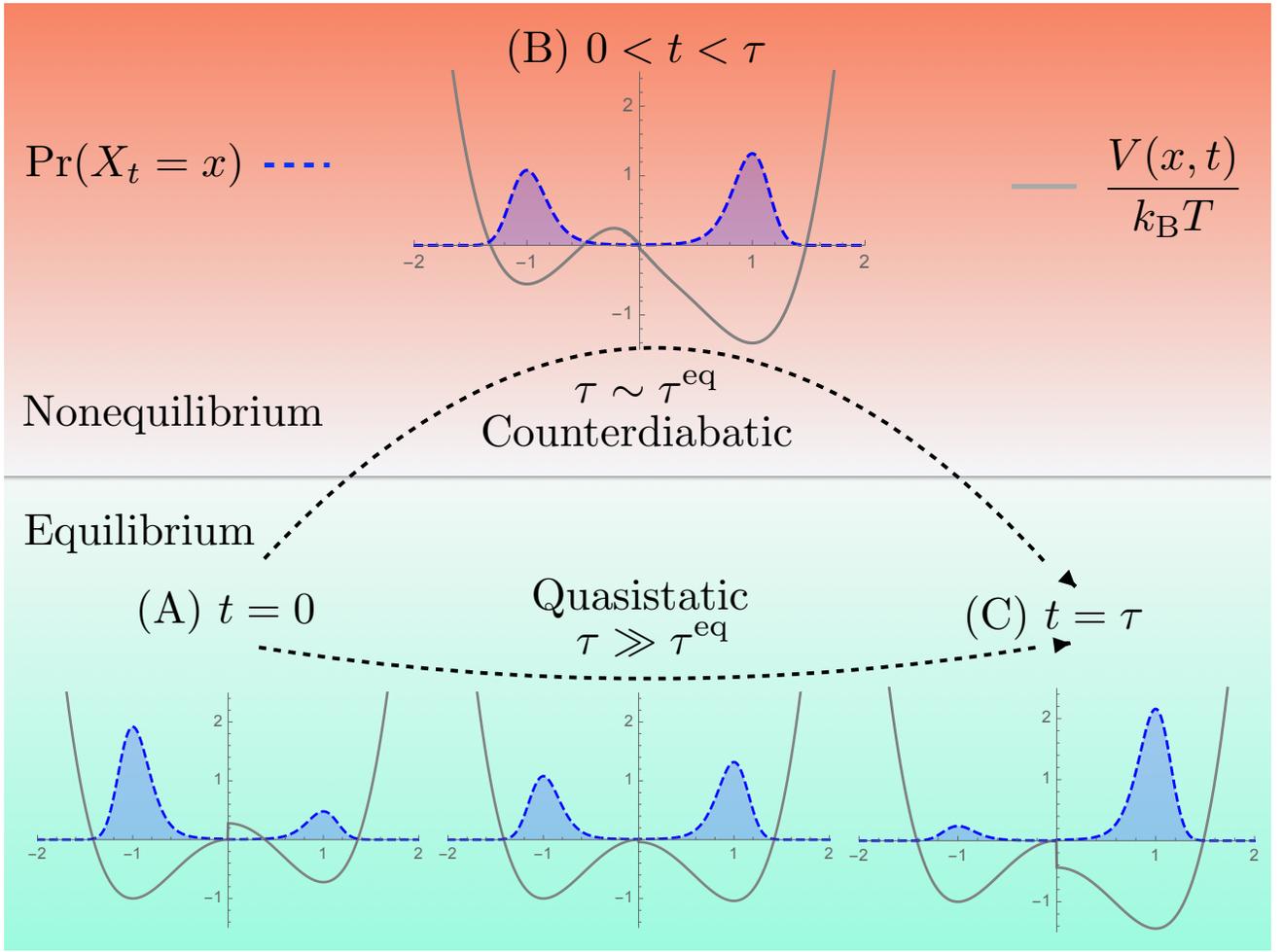}
\caption{Counterdiabatic control of the energy landscape $V(x,t)$
	(solid gray curve) at times along the interval $t \in [0,\tau]$ guides the
	probability distribution $\Pr(X_t=x)$ (dashed blue curve) along a desired
	trajectory $\Pr(X^d_t=x)$ in finite time $\tau$. The system starts in
	equilibrium in stage (A) and ends in equilibrium at stage (C), meaning that
	$V(x,0)=V^Q(x,0)$ and $V(x,\tau)=V^Q(x,\tau)$ are the quasistatic
	potentials for the initial and final distributions, respectively. However,
	at intermediate times, in stage (B), the necessary control protocol
	$V(x,t)$  required to guide the system along the desired distribution
	changes as we change the speed of the protocol. If the timescale of
	equilibration is relatively very small $\tau^\text{eq} \ll \tau$, then the
	control protocol must be in equilibrium with the desired distribution,
	such that the control potential is described by the quasistatic potential
	$V(x,t)=V^Q(x,t)$, as shown in the lower half of stage (B). Otherwise, an
	additional counterdiabatic term $V^{CD}(x,t)$ is added to the potential
	energy, which pushes the system out of equilibrium, as shown in the upper
	half of stage (B).
	}
\label{fig:CounterdiabaticVSQuasistatic}
\end{figure*}

When $\tau$ is much larger than the system's relaxation timescale
$\tau^\text{eq}$, a control protocol can be determined by assuming the system remains
approximately in equilibrium at all times: $\Pr(X_t=x) \approx
\Pr(X^\text{eq}_t=x)$. This \emph{quasistatic} (adiabatic) control protocol
is determined from the \emph{quasistatic potential}:
\begin{align}
\label{eq:VQdef}
V^Q(x,t) \equiv F^\text{eq}(t)-\kB T \ln \Pr(X^d_t=x)
  ~,
\end{align}
where the equilibrium free energy:
\begin{align}
F^\text{eq}(t) & = -\kB T \ln Z(t) \nonumber
\\ & =-\kB T \ln \int_{-\infty}^{\infty} dx e^{-V^{Q}(x,t)/\kB T}
\end{align}
is the baseline energy $E(t)$. Note that $\Pr(X^d_t)$ is the equilibrium distribution
corresponding to $V^Q(x,t)$, see Eq. (\ref{eq:BoltzmannDistFromPotential}).
In the large-$\tau$ case, the system follows this equilibrium distribution, as shown in Fig.
\ref{fig:CounterdiabaticVSQuasistatic}, and the quasistatic protocol provides the
unique solution to our control problem.
Moreover, the work invested is the change in equilibrium free energy:
\begin{align}
\langle W^Q \rangle = \Delta F^\text{eq}
  ~.
\end{align}

If, however, $\tau$ is \emph{not} much larger than $\tau^{\rm eq}$, then
evolution under the quasistatic potential $V^Q(x,t)$, defined by
Eq.~(\ref{eq:VQdef}), does not drive the system along the desired trajectory
distribution $\Pr(X_t^d)$. Rather, the actual distribution $\Pr(X_t)$
deviates from the desired distribution as the system is pushed away from
equilibrium.  Recent work \cite{Patr17a} describes how to construct a {\it
counterdiabatic protocol} that achieves the desired evolution,
$\Pr(X_t)=\Pr(X^d_t)$, for all $t \in (0,\tau)$. In this approach the
overdamped system evolves under a potential:
\begin{align}
\label{eq:Vtot}
V(x,t)=V^Q(x,t)+V^{CD}(x,t)
  ~,
\end{align}
which consists of both the quasistatic term $V^Q(x,t)$ and a
\emph{counterdiabatic potential} $V^{CD}(x,t)$, constructed to guarantee that
the actual distribution tracks the desired distribution, $\Pr(X_t) =
\Pr(X^d_t)$, as illustrated in Fig. \ref{fig:CounterdiabaticVSQuasistatic}. 

By Eq.~(\ref{eq:VQdef}) $\Pr(X^d_t)$ is the equilibrium distribution
corresponding to the quasistatic potential $V^Q(x,t)$, but it is not the
equilibrium distribution corresponding to the total potential $V(x,t)$ given by
Eq.~(\ref{eq:Vtot}). Thus, when the system evolves under the counterdiabatic
protocol, it is out of equilibrium with respect to the instantaneous potential
$V(x,t)$ at intermediate times $t \in (0,\tau)$. However, to ensure that the
system starts and ends in equilibrium, we choose $\Pr(X^d_t)$ such that
$\partial_t \Pr(X^d_t)$ vanishes at the start and end of the protocol. This way
the counterdiabatic potential vanishes at the endpoints of the
protocol: $V^{CD}(x,t=0,\tau)=0$. And so, the potential energy becomes the
quasistatic potential $V(x,t=0,\tau)=V^Q(x,t=0,\tau)$, as shown in Fig.
\ref{fig:CounterdiabaticVSQuasistatic}.

\subsection{Counterdiabatic Control of Stochastic Systems}
\label{sec:Counterdiabatic Control of Stochastic Systems}

Reference~\cite{Patr17a} showed that the counterdiabatic potential
$V^{CD}(x,t)$ is constructed from the desired distribution $\Pr(X^d_t)$ by
integrating a velocity {\it flow field} $v(x,t)$, defined shortly:
\begin{align*}
V^{CD}(x,t)=-\frac{1}{\mu} \int_{0}^xv(x',t)dx'
  ~.
\end{align*}
The lower limit of integration is set to $0$ for convenience. In fact, it may
take any value, as the physics is unchanged by the addition of an arbitrary
function $f(t)$ to the potential. The velocity flow field:
\begin{align*}
v(x,t)=\frac{\partial x}{\partial t} \biggr\rvert_{C=\text{const}} = - \frac{\partial_t C}{\partial_x C}
\end{align*}
is the velocity of constant values of the cumulative distribution function: \begin{align*}
C(x,t)=\int_{-\infty}^x \Pr(X^d_t=x') dx'
  ~.
\end{align*}
Combining results, we have, explicitly:
\begin{align}
V^{CD}(x,t)=\frac{1}{\mu} \int_{0}^x \int_{-\infty}^{x'} \frac{\partial_t\Pr(X^d_t=x'')}{\Pr(X^d_t=x')} dx'' dx' 
  ~,
\label{eq:CounterdiabaticPotential}
\end{align}
for $t \in (0,\tau)$. For $t \notin (0,\tau)$ we set $V^{CD}(x,t)=0$, hence $V(x,t)
= V^Q(x,t)$ outside of the control interval. As a result, the system begins in
the equilibrium distribution at $t=0$ and it ends (and subsequently remains) in
equilibrium at $t\ge\tau$.

Since the potential energy $V(x,t)$ consists of quasistatic and counterdiabatic
terms, we can similarly decompose the work in Eq.~(\ref{eq:Wdef}) into two
contributions:
\begin{align}
\label{eq:W}
\langle W \rangle
  & = \int_{0}^\tau dt
  \int_{-\infty}^\infty dx \Pr(X_{t}=x) \partial_t V^Q(x,t) \nonumber \\
  & \qquad + \int_{0}^\tau dt
  \int_{-\infty}^\infty dx \Pr(X_{t}=x) \partial_t V^{CD}(x,t) \nonumber \\
  & = \langle W^Q \rangle + \langle W^{CD} \rangle  \nonumber \\
  & = \Delta F^\text{eq}+\langle W^{CD} \rangle
  ~.
\end{align}
The first term $\langle W^Q \rangle$ is the amount of work that would be
performed \emph{if} the protocol were executed quasistatically, i.e.,
reversibly. This \emph{quasistatic work} is simply the change in equilibrium
free energy, as follows by direct substitution of Eq.~(\ref{eq:VQdef}) into the
first line above. This contribution depends only on the initial and final
potential and not on either (i) the sequence of intermediate distributions or
(ii) the duration of the protocol.

The second contribution $\langle W^{CD} \rangle$ is the \emph{counterdiabatic
work}, and it is proportional to the global entropy production $\langle \Sigma
\rangle$. Specifically, when the system begins and ends in equilibrium we
have~\cite{Jarzynski2011}:
\begin{equation}
\label{eq:TSigma}
T\langle \Sigma \rangle  = \langle W \rangle -\Delta F^\text{eq} 
  = \langle W^{CD} \rangle
  ~,
\end{equation}
where $\langle \Sigma\rangle \ge 0$ quantifies the net change in the entropy of
the system and its thermal surroundings.

In Eq.~(\ref{eq:W}), the quasistatic work is fixed and the counterdiabatic work
gives the path-dependent dissipated work required to complete the
transformation in finite time. Thus, all dependence on intermediate details is
captured by $\langle W^{CD} \rangle$. This quantity is our principal focus and,
as we now show, it scales particularly simply with system size and computation
time.

We note that Eqs.~(\ref{eq:W}) and (\ref{eq:TSigma}), along with the inequality
$\langle\Sigma\rangle \ge 0$, generalize to transformations between
nonequilibrium states, with $\Delta F^\text{eq}$ replaced by the recoverable
nonequilibrium free energy, $\Delta F^\text{neq}$; see Refs.~\cite{Parr15a,
Espo11, Taka10a} for details. We will use this generalized result in
Sec.~\ref{sec:CDErasure} when discussing counterdiabatic erasure.

\subsection{System-Size and Computation-Rate Dependence}
\label{sec: Size and Computation Rate Dependence}

A protocol's \emph{duration} $\tau$ is the time over which the Hamiltonian
varies. For our one-dimensional system, we define a characteristic \emph{system
length} $L$ reflecting the extent of the desired probability distribution's
support. Since we wish to capture only the distribution's bulk and not the
support's absolute extent, there are many ways to define this length. A
candidate is the initial variance:
\begin{align*}
L= \sqrt{\int_{-\infty}^{\infty}dx \Pr(X^d_0=x)x^2
  - \left(\int_{-\infty}^{\infty}dx \Pr(X^d_0=x)x \right)^2}
  .
\end{align*}
The particular form is somewhat arbitrary. All we ask is that $L$ scale
appropriately when transforming the distribution. With these definitions in
hand, we can analyze how the protocol and dissipation change under rescalings.

Consider the probability trajectory $\{\Pr(X^d_t=x): ~t \in (0,\tau)\}$ and a
system of length $L$, yielding the control protocol
$V(x,t)=V^Q(x,t)+V^{CD}(x,t)$. To preserve the probability trajectory shape
while changing the duration to $\tau'$ and length to $L'$, we introduce a new
desired trajectory:
\begin{align*}
\Pr(X_t^{d \prime}=x)=\Pr(X^d_{\tau t/\tau'}=Lx/L')\frac{L}{L'}
  ~.
\end{align*}
This stretches the original distribution's support by a factor $L'/L$ and increases the
computation rate by a factor $\tau/\tau'$.

In the expression for the resulting counterdiabatic control protocol:
\begin{align*}
V'(x,t)=V'^Q(x,t)+V'^{CD}(x,t)
  ~,
\end{align*}
we define a new quasistatic potential as the similarly-scaled version of the
original:
\begin{align*}
V^{Q\prime}(x,t)=V^Q(Lx/L',\tau t / \tau')
  ~.
\end{align*}
The associated equilibrium free energy is expressed in terms of the original free energy:
\begin{eqnarray}
F^{\text{eq}\prime}(t) &=& -\kB T \ln Z'(t) \nonumber \\
  &=& -\kB T \ln \int_{-\infty}^{\infty} dx e^{-V^{Q\prime}(x,t)/\kB T} \nonumber \\
  &=& -\kB T \ln \int_{-\infty}^{\infty} dx'
  \frac{L'}{L}e^{-V^{Q}(x',\tau t/\tau')/\kB T} \nonumber \\
  \label{eq:Fprime}
  &=& \kB T \ln \frac{L}{L'} +F^\text{eq}(\tau t/\tau')
  ~,
\end{eqnarray}
where the third line comes from substituting $x = x' L'/L$.
Equation (\ref{eq:Fprime}) implies:
\begin{equation*}
\Delta F^{\text{eq} \prime}  = F^{\text{eq} \prime}(\tau')-F^{\text{eq} \prime}(0)  = \Delta F^\text{eq} 
  ~,
\end{equation*}
hence the quasistatic work is the same for protocols with different
durations and lengths:
\begin{align*}
\langle W^{Q \prime} \rangle=\langle W^{Q} \rangle
  ~.
\end{align*}

The counterdiabatic contributions, however, yield meaningful differences when
changing system length or protocol duration. Substituting the rescaled
probability trajectory into the expression for counterdiabatic potential in Eq.
(\ref{eq:CounterdiabaticPotential}), we find:
\begin{align*}
& V^{CD \prime}(x,t) = \frac{1}{\mu} \int_{0}^x \int_{-\infty}^{x'}
	\frac{\partial_t\Pr(X^{d \prime}_t=x'')}{\Pr(X^{d\prime}_t=x')} dx'' dx' \\
  & \quad = \frac{1}{\mu} \int_{0}^x \int_{-\infty}^{x'}
  \frac{\partial_t \Pr(X^{d}_{\tau t/ \tau'}=Lx''/L')}
  {\Pr(X^{d}_{\tau t/ \tau'}=Lx'/L')} dx'' dx' \\
  & \quad = \frac{1}{\mu} \frac{L^{\prime 2}}{L^2} \int_{0}^{Lx/L'}
  \int_{-\infty}^{x'''}
  \frac{\partial_{t'}\Pr(X^{d}_{t'}=x'''') \partial_t t'}
  {\Pr(X^{d}_{t'}=x''')} dx'''' dx''' \\
  & \quad = \frac{\tau L^{\prime 2}}{ \tau' L^2}V^{CD}(Lx/L',\tau t/ \tau')
  ~,
\end{align*}
using the substitutions $t' = \tau t /\tau'$,
$x'''=Lx'/L'$, and $x''''=Lx''/L'$. Thus, the counterdiabatic potential scales
as the square of the length of the information storage device and as the
inverse of the protocol duration. Equivalently, the additional nonequilibrium
force $F^{CD}(x,t)=-\partial_x V^{CD}(x,t)$ applied to the system scales
as the computation rate and square of the system size.

For the counterdiabatic work we similarly find:
\begin{align*}
& \langle W^{CD \prime} \rangle = \int_{0}^{\tau'} \!\!\!\! dt
  \int_{-\infty}^\infty \!\!\!\! dx \Pr(X^{d\prime}_{t}=x)
  \partial_t V^{CD \prime}(x,t) \\
  & = \frac{\tau L^{\prime 2}}{ \tau' L^2} \int_{0}^{\tau'} \!\!\!\! dt
  \int_{-\infty}^\infty \!\!\!\! dx \frac{L}{L'}
  \Pr \left(X^{d}_{\tfrac{\tau}{\tau'} t } \!\! =  \!\! \frac{L}{L'}x \right)
  \partial_t V^{CD}\left(\frac{L}{L'}x, \frac{\tau}{\tau'} t \right) \\
  & = \frac{\tau L^{\prime 2}}{ \tau' L^2} \int_{0}^{\tau} \frac{\tau'}{\tau}dt' \int_{-\infty}^\infty \frac{L'}{L}dx' \frac{L}{L'}\Pr(X^{d}_{t'}=x') \partial_t V^{CD}(x',t') \\
  & = \frac{\tau L^{\prime 2}}{ \tau' L^2} \frac{\tau'}{\tau}
  (\partial_t t')\int_{0}^{\tau} \!\!\!\! dt'
  \int_{-\infty}^\infty  \!\!\!\! dx' \Pr(X^{d}_{t'}=x')
  \partial_{t'} V^{CD}(x',t') \\
  & = \frac{\tau L^{\prime 2}}{ \tau' L^2} \langle W^{CD} \rangle
  ~.
\end{align*}
And so, too, the dissipated counterdiabatic work scales as system
length squared and linearly with computation rate. This work, in turn, is
proportional to the entropy production, so we find that the entropy production
obeys a similar scaling:
\begin{align*}
\langle \Sigma' \rangle
  & = \frac{\langle W^{CD \prime} \rangle}{T} \\
  & =  \frac{\tau L^{\prime 2}}{ \tau' L^2}\langle \Sigma \rangle
  ~.
\end{align*}

\subsection{Efficient Protocols}
\label{sec:Efficient Protocols}

When changing the protocol duration $\tau \rightarrow \tau'$ and
system length $L \rightarrow L'$ of a desired distribution trajectory
$\{\Pr(X^d_t)\}$, the counterdiabatic control becomes:
\begin{align*}
V'(x,t) = V^Q \left(\frac{L}{L'}x,\frac{\tau}{\tau'}t \right)
  + \frac{\tau L^{\prime 2}}{ \tau' L^2}
  V^{CD} \left(\frac{L}{L'}x,\frac{\tau}{\tau'}t \right)
  ~,
\end{align*}
where $V^Q(x,t)$  and $V^{CD}(x,t)$ are the original quasistatic and  counterdiabatic potential energies.
This leads to the work investment:
\begin{align*}
\langle W'\rangle =\Delta F^\text{eq}+\frac{\tau L^{\prime 2}}{ \tau' L^2} \langle W^{CD} \rangle
  ~,
\end{align*}
where $\Delta F^\text{eq}$ is the original change in free energy and $\langle W^{CD} \rangle$ is the original nonequilibrium addition to work.

Since protocols are uniquely determined by the distribution trajectory, the
above scaling relations apply directly to maximally efficient computations as
well. Since a computation maps an initial equilibrium distribution
$\Pr(X_0)$ to a final one $\Pr(X_\tau)$, there are many compatible
distribution trajectories that evolve continuously from the initial to the
final distribution. A minimally dissipative distribution trajectory
$\Pr(X_{t,\text{min}})$ has a corresponding
$V_\text{min}(x,t)=V^Q_\text{min}(x,t)+V^{CD}_\text{min}(x,t)$ that yields the
minimum work:
\begin{align*}
\langle W^{CD}\rangle_\text{min} =
  \min\{\langle W^{CD}\rangle : \Pr(X_{0,\tau}^d)=\Pr(X_{0,\tau}) \}
  ~.
\end{align*}
Since quasistatic work is identical for all such protocols, up to an instantly
recoverable additional energy, this condition also minimizes invested work.

Changing protocol duration $\tau \rightarrow \tau'$ and initial and final system
length---viz., $\Pr(X'_0=x)=\Pr(X_0=Lx/L')$ and
$\Pr(X'_{\tau'}=x)=\Pr(X_\tau=Lx/L')$---we can determine how the minimally
dissipative distribution trajectory changes, as well the minimum dissipation. A
natural guess for the minimally dissipative trajectory is to take the scaled
minimal distribution:
\begin{align*}
\Pr(X^{\prime}_{t}=x)=\Pr(X_{\tau t/\tau',\text{min}}=Lx/L')\frac{L}{L'}
  ~,
\end{align*}
which satisfies:
\begin{align*}
\langle W^{CD \prime} \rangle =\frac{\tau L^{\prime 2}}{ \tau' L^2} \langle W^{CD} \rangle_\text{min}
  ~.
\end{align*}
(See Sec.~\ref{sec: Size and Computation Rate Dependence}.)

If this proposed trajectory is not minimally dissipative, then there is another
trajectory $\{\Pr(X^\prime_{t,\text{min}})\}$ that dissipates work $ \langle
W^{CD \prime} \rangle_\text{min} < \langle W^{CD \prime} \rangle$. However, if
this were the case, then we could reverse the duration and size scalings $\tau'
\rightarrow \tau$ and $L' \rightarrow L$ on that trajectory to generate the
dissipation:
\begin{align*}
\frac{\tau' L^2 }{\tau L^{\prime 2}}\langle W^{CD \prime} \rangle_\text{min}
 & <\frac{\tau' L^2 }{\tau L^{\prime 2}}\langle W^{CD \prime} \rangle \\
 & =\frac{\tau' L^2 }{\tau L^{\prime 2}}
 \frac{\tau L^{\prime 2}}{ \tau' L^2}
 \langle W^{CD} \rangle_\text{min} \\
 & =\langle W^{CD} \rangle_\text{min}
  ~.
\end{align*}
This is a contradiction, since it states that it is possible to dissipate
less than the minimal dissipation for the original computation that evolves the
distribution between $\Pr(X_0)$ and $\Pr(X_\tau)$. We conclude that
the spatially- and temporally-scaled minimally dissipative distribution
trajectories are themselves minimally dissipative.

While this does not provide a method for discovering the minimally dissipative
protocol, paralleling other approaches in the restricted linear regime
\cite{Zulk14a,Lahi16a}, it shows how to achieve maximal efficiency given other
constraints on space and time, if one finds the minimally dissipative protocol
in one setting. Moreover, it gives the temporal scaling of the minimally
dissipative control protocol:
\begin{align*}
V^\prime_\text{min}(x,t)
  & = V^Q_\text{min}(Lx/L',\tau t/\tau') \\
  & \qquad + \frac{\tau L^{\prime 2}}{ \tau' L^2} V^{CD}_\text{min}(Lx/L',\tau t/\tau')
  ~,
\end{align*}
and of the minimum work production:
\begin{align*}
\langle W^\prime \rangle_\text{min}
  = \Delta F^\text{eq}
  +\frac{\tau L^{\prime 2}}{ \tau' L^2}\langle W^{CD}\rangle_\text{min}
  ~.
\end{align*}

This matches independent analyses on the scaling of dissipated work for optimal
control \cite{Zulk14a}. In contrast, as we see, the present results apply more
generally: without restricting control parameters---all potential landscapes
are allowed---and beyond linear response. This perhaps explains the puzzle that
the results derived assuming linear response \cite{Zulk12a,Zulk14a} appeared to work outside of those
constraints. Moreover and constructively, counterdiabatic protocols allow a
control engineer to specify exact start and ending distributions. The latter
must be inferred from dynamics in other treatments.

In short, counterdiabatic control of Fokker-Planck dynamics in one dimension
gives precise control over distributions and yields constructive methods for
designing control protocols. The resulting energetic costs depend simply on
overall system temporal and spatial scales, revealing a tradeoff beyond that
between a computation's information processing and energy cost.
 
\section{Counterdiabatic Erasure}
\label{sec:CDErasure}

We now apply the counterdiabatic approach to the paradigmatic example of
erasing a bit of information in a metastable system. The analysis exposes new
elements in the resource tradeoffs that arise in thermodynamic computing.

\subsection{Nonequilibrium Information Storage}
\label{sec:Nonequilibrium Information Storage}

The ability to quickly shift probability distributions in one-dimensional
nonlinear Langevin systems gives a physical way to implement fast logical
operations. For instance, erasure is an essential part of most computations and
can be achieved by controlling a double-well potential landscape
\cite{Berut12,Jun14a}. Landauer stated that erasure requires dissipating at
least $\kB T \ln 2$ of work---a cost deriving from the microstate space
contraction induced by the logically irreversible operation \cite{Land61a}.
This bound is indeed achievable in the present setting, but only in the limit
of quasistatic operations, where zero entropy is produced globally. That is, it
is achievable only in infinite time. For finite-time processes, there is
additional dissipation and, thus, additional work required for erasing a bit of
information \cite{Berut12,Jun14a, Zulk14a}. We will now derive the same
additional cost for finite-time erasure but, departing from prior treatments,
we determine the initial and final distributions and thus exactly specify the
fidelity of erasure, instead of merely recreating it. This provides a detailed
analysis of thermodynamic resources for a given accuracy level of information
processing. Additionally, we will design a protocol that gives perfect erasure
in finite time and at finite cost.

The expression for the counterdiabatic potential Eq.
(\ref{eq:CounterdiabaticPotential}) specifies how to design a protocol $V(x,t)$
that maintains the distribution $\Pr(X_t)$ exactly in a desired distribution
$\Pr(X^d_t)$ over the interval $t \in (0,\tau)$. However, we must also consider
how to use the map to informational states $c:\mathcal{X} \rightarrow
\mathcal{Y}$ to perform useful and robust computation. One strategy is to
design the energy landscape such that physical states $x \in \mathcal{X}$ in
one informational state $y \in \mathcal{Y}$ rarely transition to different
informational states $y' \neq y$. This allows the information processing device
to remain in a passive ``default'' state while retaining the information of its
computation for long times, regardless of the outcome. In contrast, if a
computation is designed such that the equilibrium distribution
$\Pr(X^\text{eq}_t)$ exactly matches the desired distribution $\Pr(X^d_t)$
after the computation, with $t>\tau$, then the energy landscape is given by
$V(x,t)=F^\text{eq}(t)-\kB T \ln \Pr(X^d_t)$ for $t \geq \tau$.  The potential
energy characterizes the external configuration of our memory storage device,
and it is through control of this external configuration that $V(x,t)$ has time
dependence.

Thus, the relevant information about the computation is stored in the external
configuration of the memory device $V(x,t)$ (our control), rather than the
actual physical distribution $\Pr(X_t)$. By choosing a default energy landscape
that stores metastable physical distributions, the computational device can
robustly store information without explicitly encoding the outcome distribution
of the computation in the energy landscape and thus the external configuration.
With metastable information, a memory device can store information about its
input that is not explicitly encoded in the control parameters.

\begin{figure}[tbp]
\centering
\includegraphics[width=\columnwidth]{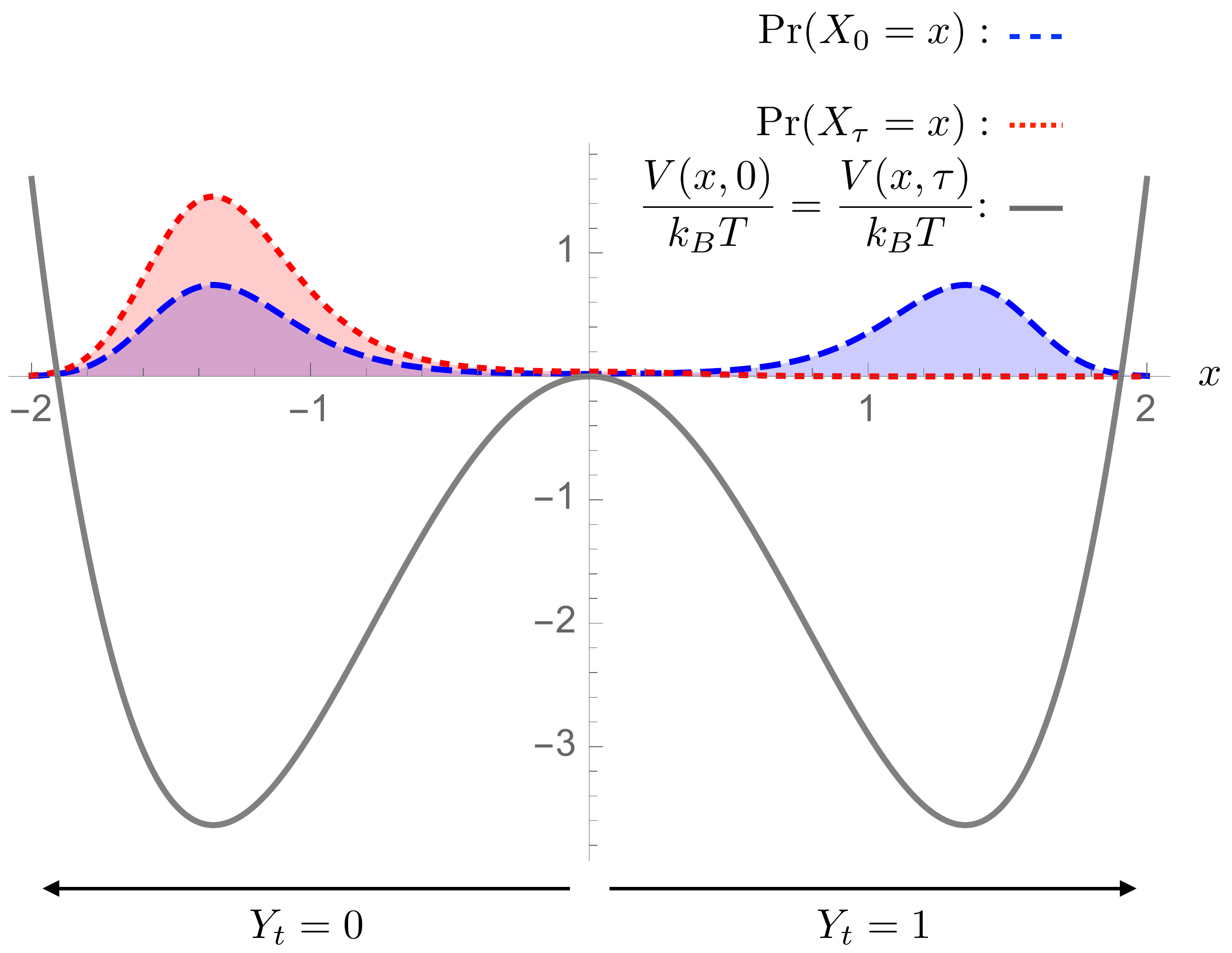}
\caption{Default energy landscape: A double-well potential that stores many
	different distributions $\Pr(Y_t)$ over the informational states $Y_t \in
	\{0,1\}$. For protocols that process the information in the distributions
	over the times $t \in (0,\tau)$, the energy landscape is set to be the same
	at the beginning and end, shown by the gray curve $V(x,0)=V(x,\tau)$. The
	equilibrium distribution, delineated by the dashed blue curve, gives equal
	probabilities for the informational states: $\Pr(Y_0=0)=\Pr(Y_0=1)=1/2$.
	This is the initial distribution for the system $\Pr(X_0)$ in this case.
	It stores $\H[\Pr(Y_0)] = 1$ bit of information, where $\H[Z]$ is the
	Shannon information of random variable $Z$ \cite{Cove06a}. The red curve
	$\Pr(X_\tau=x)$ gives unit probability of informational state $0$
	($\Pr(Y_\tau=0)=1$) and is the distribution of the system after an
	effective erasure protocol. Its Shannon information vanishes and so the
	initial and final distributions represent bit erasure. The final
	distribution $\Pr(X_\tau)$ is out of equilibrium, but the energy barrier
	between the two informational states keeps it nearly fixed for short times.
	This distribution, as well as many other nonequilibrium distributions, are
	metastable, and will eventually, slowly relax to equilibrium.
	}
\label{fig:MetastableDistributions}
\end{figure}

To experimentally test Landauer's prediction \cite{Land61a}, Ref.
\cite{Jun14a} employed a protocol that starts and ends in a symmetric
double-well potential, where each well is interpreted as a distinct mesoscopic
informational state: $Y_t = 0$ or $Y_t = 1$. Such a potential stores
informational states determined by the probability $\Pr(Y_t=0)$ to realize the
informational state $0$. Following this setup, if we start and end in a
symmetric bistable potential:
\begin{align}
V(x,0)=V(x,\tau)=\alpha x^4- \beta x^2
  ~,
\label{eq:DefaultPotential}
\end{align}
then, at a temperature $T$, the equilibrium distribution:
\begin{align}
\label{eq:px}
p(x) \equiv \Pr(X_{\{0,\tau\}}^\text{eq}=x)=\frac{e^{-V(x,0)/\kB T}}{Z}
  ~,
\end{align}
is bimodal; see the blue dashed curve in Fig.
\ref{fig:MetastableDistributions}. 
While this distribution is exactly stationary (when the potential is held fixed),
we can construct other
distributions that are (temporarily) effectively stationary, such as that given
by the dotted red curve shown in Fig. \ref{fig:MetastableDistributions}. This
has the same shape as the equilibrium distribution over the interval
$(-\infty,0)$, but vanishes outside. By specifying a time-dependent \emph{bit bias}
probability $\Pr(Y_t=0)=b(t)$, we fully specify a metastable physical distribution \cite{Hang90a}:
\begin{align}
\label{eq:Pmet}
\Pr(X^\text{met}_t=x) =
\begin{cases}
	p(x)\cdot 2b(t) & \text{ if } x\leq 0 \\
	p(x)\cdot 2(1-b(t)) & \text{ if } x> 0
\end{cases}
	~.
\end{align}
We take this distribution to be our desired distribution $\Pr(X^d_t)$,
which in turn defines the quasistatic potential $V^Q(x,t)$.

Figure \ref{fig:MetastableDistributions} shows the metastable distributions
before (blue dashed curve) and after (red dotted curve) an erasure protocol,
where the initial distribution is unbiased $b(0)= 1/2$, and the final
distribution is totally biased $b(\tau)=1$. The energy barrier between
informational states $0$ and $1$ inhibits large probability flow between the
two local equilibria. That is, these distributions relax to a global
equilibrium very slowly, depending on barrier height relative to $\kB T$
\cite{Hang90a}. Thus, these metastable states robustly store nonequilibrium
informational states and provide a basis for information processing by a
controlled double-well potential.

\begin{figure*}[tbp]
\centering
\includegraphics[width=2\columnwidth]{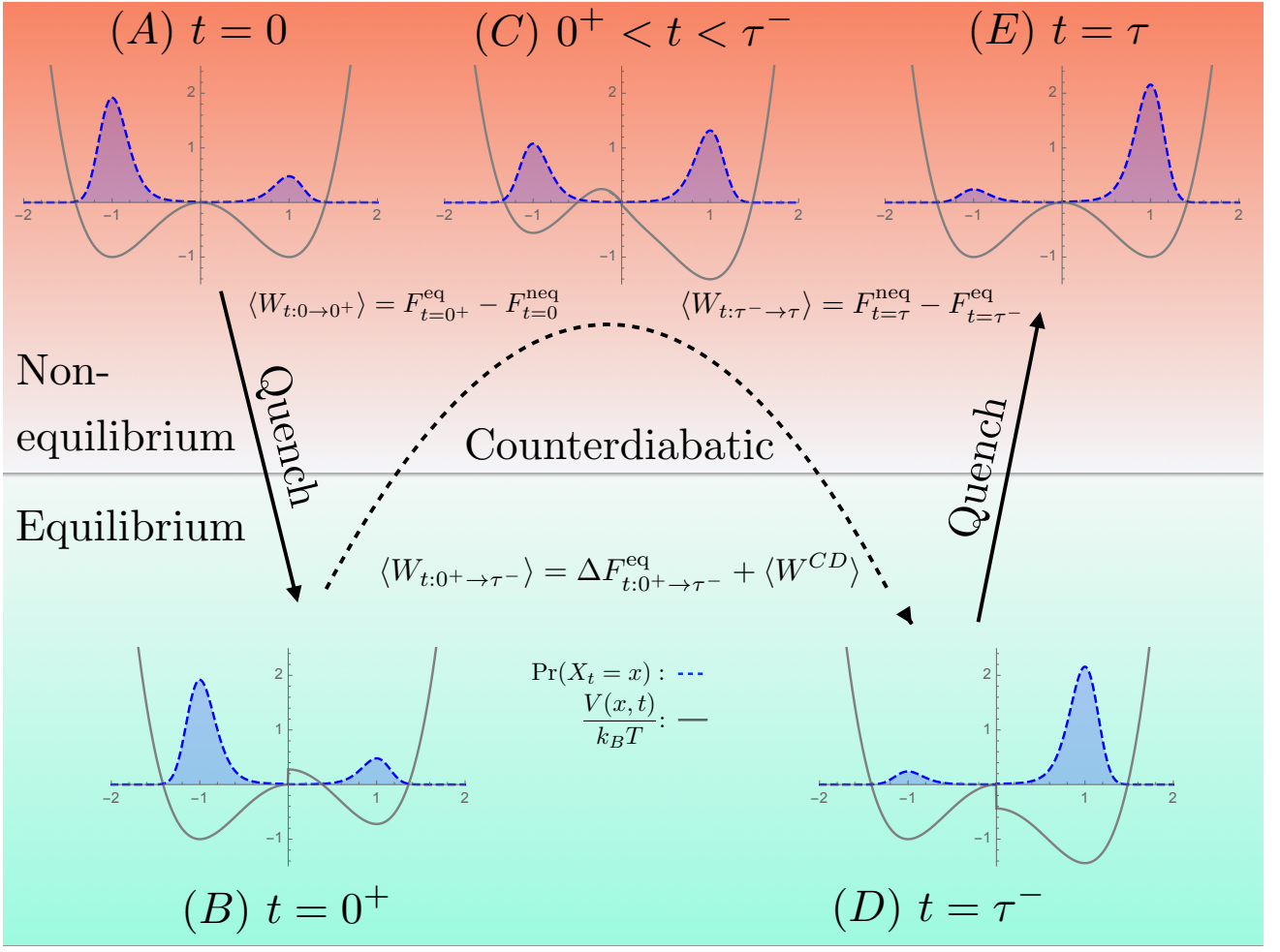}
\caption{Counterdiabatic information processing in three steps: Distribution
	$\Pr(X_t=x)$ shown with blue dashed curves and energy landscapes $V(x,t)$
	shown by the gray curves. First, the information landscape is
	instantaneously changed to fit the starting distribution from stage (A) to
	stage (B). These share the same distribution but have different energy
	landscapes. Second, the counterdiabatic protocol is applied to take the
	system from the equilibrium distribution at stage (B) to that shown
	in stage (D), passing through nonequilibrium distributions driven from
	equilibrium by the counterdiabatic potential; such as that shown in
	stage (C). Third, the last quench step instantaneously takes the system
	from equilibrium stage (D) to the nonequilibrium metastable stage (E).
	All three transitions are labeled with the associated work investment.
	}
\label{fig:CDInfoProcessing}
\end{figure*}

\subsection{Counterdiabatic Information Processing}
\label{sec:Counterdiatic Information Processing}

We now consider how to use counterdiabatic driving to steer the system through
a sequence of metastable states specified by a given time-dependent bit bias
$b(t)$, with the symmetric initial and final energy landscape
of Eq.~(\ref{eq:DefaultPotential}). Despite the symmetric initial and final configurations of the memory device, this modified counterdiabatic control allows for $b(0) \ne 1/2$ and $b(\tau) \ne 1/2$.

Since the initial and final metastable states are out of equilibrium with
respect to the symmetric potential $V(x,0)=V(x,\tau)=\alpha x^4-\beta x^2$
(Fig.~\ref{fig:MetastableDistributions}), we must modify the counterdiabatic
protocol described in Sec.~\ref{sec:Counterdiabatic Control of Stochastic
Systems}, which was developed for transitions between initial and final
equilibrium distributions. Two additional steps are needed, each a quench, as
shown in Fig. \ref{fig:CDInfoProcessing}. ({\it Quench} here means a nearly
instantaneous change in the Hamiltonian \cite{Kawa07}, as opposed to a nearly
instantaneous change in temperature, as often intended.) These quenches are
added to make the quasistatic potential $V^Q(x,t)$ match the equilibrium
distribution of the desired metastable distribution $\Pr(X^\text{met}_t)$ over
the open time interval $t \in (0 ,\tau)$.

Specifically, for $t \in (0,\tau)$ we set:
\begin{align*}
V^Q(x,t)=F^\text{eq}(t)-\kB T \ln \Pr(X^\text{met}_t=x)
  ~.
\end{align*}
Hence, at $t=0$ the energy landscape undergoes a quench from the symmetric
potential $V(x,0)$ to the asymmetric potential $V^Q(x,0)$. We then add the
counterdiabatic term:
\begin{align*}
V^{CD}(x,t) =\frac{1}{\mu} \int_{0}^x \int_{-\infty}^{x'} \frac{\partial_t\Pr(X^\text{met}_t=x'')}{\Pr(X^\text{met}_t=x')}dx'' dx',
\end{align*}
such that the overall potential becomes:
\begin{align*}
V(x,t)=V^{Q}(x,t)+V^{CD}(x,t)
  ~.
\end{align*}
For $t\in(0,\tau)$ the system evolves through the desired sequence
$\Pr(X^\text{met}_t=x)$, corresponding to the equilibrium states of $V^Q(x,t)$.
At the end of the protocol the system undergoes another quench, from the
asymmetric potential $V^Q(x,\tau^-)$ to the symmetric potential $V(x,\tau)$. In
this way, we drive the system through a sequence of metastable distributions
with precise control of the bit bias $b(t)$.


Although the protocol just described pertains to the specific case of a double
well, the procedure of quenching, controlling counterdiabatically, and then
quenching again is a general technique for evolving between nonequilibrium
distributions in finite time. For such a computation, the total work simplifies
to the net change in nonequilibrium free energy plus the counterdiabatic work:
\begin{align}
\langle W \rangle = \Delta F^\text{neq}+\langle W^{CD}\rangle
  ~,
\label{eq:totalWork}
\end{align}
as shown in Fig. \ref{fig:CDInfoProcessing}. The change in nonequilibrium free
energy is given by the sum of the quasistatic work and the quenching work
\cite{Espo11,Parr15a}. For the metastable distributions we discussed, where
each informational state contributes the same local free energy, $\Delta
F^\text{neq}$ reduces to the  change in the Shannon entropy of the information
variable \cite{Parr15a}:
\begin{align}
\label{eq:dFneq}
\Delta F^\text{neq}
  & = \kB T \ln 2 \left( \H[Y_0]-\H[Y_\tau] \right)
  ~.
\end{align}
Since $\langle W^{CD}\rangle = T \langle \Sigma \rangle \ge 0$ (see
Sec.~\ref{sec:Counterdiabatic Control of Stochastic Systems}),
Eqs.~(\ref{eq:totalWork}) and (\ref{eq:dFneq}) lead immediately to a
generalized form of Landauer's bound:
\begin{align}
\langle W\rangle \ge \kB T \ln 2 \left( \H[Y_0]-\H[Y_\tau] \right)
  ~,
\end{align}
which takes on the familiar form, $\langle W \rangle \geq \kB T \ln 2$, when
starting with fully randomized bits, $b(0)=1/2$ and when the operation's
fidelity is perfect, $b(\tau)=1$. As we shall see, while the Landauer bound
cannot be achieved in finite time, perfect fidelity can be achieved in finite
time with finite work.

The amount of entropy produced, $\langle \Sigma \rangle = \langle W^{CD}
\rangle/T$, reflects the additional cost beyond Landauer's bound of implementing
a computation in finite time. For metastable erasure in a symmetric double
well, this additional cost depends on duration, system length scale, bit bias
difference, and information lifetime---a measure of information storage
robustness. We have already seen that the value of $\langle \Sigma \rangle$
scales as the inverse of the protocol duration $\tau$ and the square of the
system characteristic length scale $L$ (Sec.~\ref{sec: Size and Computation
Rate Dependence}). We now show how bit bias difference and information lifetime
lead to additional energy costs.

Metastability simplifies the expression for the counterdiabatic potential,
leading to simple relationships between the work, bit bias difference, and
robustness of information storage. As shown in Appendix \ref{app:Symmetric
Metastable Erasure}, the counterdiabatic potential can be expressed as a
product of a piecewise-continuous function and a function that depends only on
the equilibrium distribution:
\begin{align}
V^{CD}(x,t)
  & = h(x) \times \begin{cases}
  - \partial_t \ln b(t)     & \text{ if } x\leq 0 \\
  - \partial_t \ln (1-b(t)) & \text{ if } x>0
\end{cases}
  ,
\label{eq:MetastableVCD}
\end{align}
where:
\begin{align*}
h(x) = \frac{1}{\mu}\int_0^{|x|}dx'
  \frac{1}{p(x')} \int_{-\infty}^{-|x'|}dx''p(x'')
\end{align*}
and $p(x)=\Pr(X_0^\text{eq}=x)$ is the equilibrium distribution for the
symmetric, bistable potential of Eq.~(\ref{eq:px}). This result allows us to
design protocols for evolving a metastable distribution from an initial bit
bias $b(0)=b_i$ to any final bit bias $b(\tau)=b_f$, given a bistable
potential. For instance, the choices $b_i=1/2$ and $b_f=1$ correspond to
perfect erasure, where the system ends entirely in the left well.

\subsection{Tradeoffs in Metastable Symmetric Erasure}
\label{sec:Tradeoffs In Metastable Symmetric Erasure}

As discussed above, the equilibrium distribution $p(x)$ and bit bias $b(t)$
determine both the desired metastable distribution trajectory of
Eq.~(\ref{eq:Pmet}) and the counterdiabatic potential of
Eq.~(\ref{eq:MetastableVCD}) that generates this evolution. In
Appendix~\ref{app:Symmetric Metastable Erasure} we show that the functions
$p(x)$ and $b(t)$ are multiplicatively separable in the expression for
counterdiabatic work. Specifically:
\begin{align*}
\langle W^{CD} \rangle =f_1[p(\cdot)]f_2[b(\cdot)]
  ~,
\end{align*}
where:
\begin{align}
f_1[p(\cdot)] &= \frac{1}{\mu}
  \int_0^\infty dx \,p(x) \int_0^{x}dx' \frac{1}{p(x')}
  \int_{-\infty}^{-x'}dx''p(x'') \nonumber\\
f_2[b(\cdot)] &= \int_{0}^1 dt \frac{ (\partial_t b(t))^2}{b(t)-b(t)^2}
  ~.
\label{fig:fidelityFunctional}
\end{align}
This separability entails additional tradeoffs between dissipation, bit bias
difference, and information lifetime.

Functional $f_1$ depends on the equilibrium distribution $p(x)$ which in turn
is determined by the bistable potential $V(x,0)$. The shape of this potential
(e.g., the height of the barrier relative to the left and right minima)
determines the expected ``lifetime'' of a stored bit, in the absence of
external driving. Thus, $f_1$ contains all the dependence of the
counterdiabatic work on information storage robustness.

Functional $f_2$ depends on the bit bias trajectory $b(t)$. One can now
entertain a variety of bias trajectories, using this functional to determine
how the counterdiabatic work changes. However, note that one must restrict to
paths for which the initial and final time-derivative vanishes $\partial_t
b(t)\rvert_{t \in \{0, \tau\}}=0$, since $\partial_t \Pr(X_t=x)_{t \in \{0, \tau\}}$ must
vanish for the counterdiabatic potential itself to be zero initially and
finally.


Note too that $f_1$ and $f_2$ contain the system length and protocol duration dependence, respectively. If we
rescale the system spatially and the protocol temporally, we obtain the new equilibrium distribution and bias trajectory:
\begin{align*}
p'(x) & = \frac{L}{L'}p(Lx/L') ~\text{and}\\
b'(t) & = b(\tau t/\tau')
  ~.
\end{align*}
Plugging these in, we find the new functionals:
\begin{align*}
f^{1}[p'(\cdot)] & = \frac{L^{\prime 2}}{L^2}f^{1}[p(\cdot)] ~\text{and}\\
  f^{2}[b'(\cdot)] & = \frac{\tau}{\tau'}f^{2}[b(\cdot)]
  ~.
\end{align*}

To further separate dependencies, we introduce unitless functionals of both 
bias and the default equilibrium distribution:
\begin{align*}
F_1[p(\cdot)] & = f_1[p(\cdot)]/L^2 ~\text{and}\\
F_2[b(\cdot)] & = f_2[b(\cdot)]\tau
  ~.
\end{align*}
$F_2$ captures the difference between initial and final bias without dependence
on computation rate. $F_1$ captures the depth between left and right wells
without dependence on the spatial scale, which also affects how long a bistable
system can robustly store information. In short, the counterdiabatic work is
expressed as the product of four terms:
\begin{align*}
\langle W^{CD} \rangle= \frac{L^2}{\tau}F_1[p(\cdot)]F_2[b(\cdot)]
  ~.
\end{align*}
Since $F_1$ and $F_2$ are independent of duration and system length, this
implies a five-way tradeoff between the main dependencies of computation:
dissipation, duration, length, $F_1$, and $F_2$. We next study
how $F_1$ and $F_2$ depend on properties of the erasure protocol,
leading to a practical consequence.

\subsection{Perfect Erasure in Finite Time with Finite Work}
\label{sec:Perfect Erasure In Finite Time and Finite Work}

\begin{figure}[tbp]
\centering
\includegraphics[width=\columnwidth]{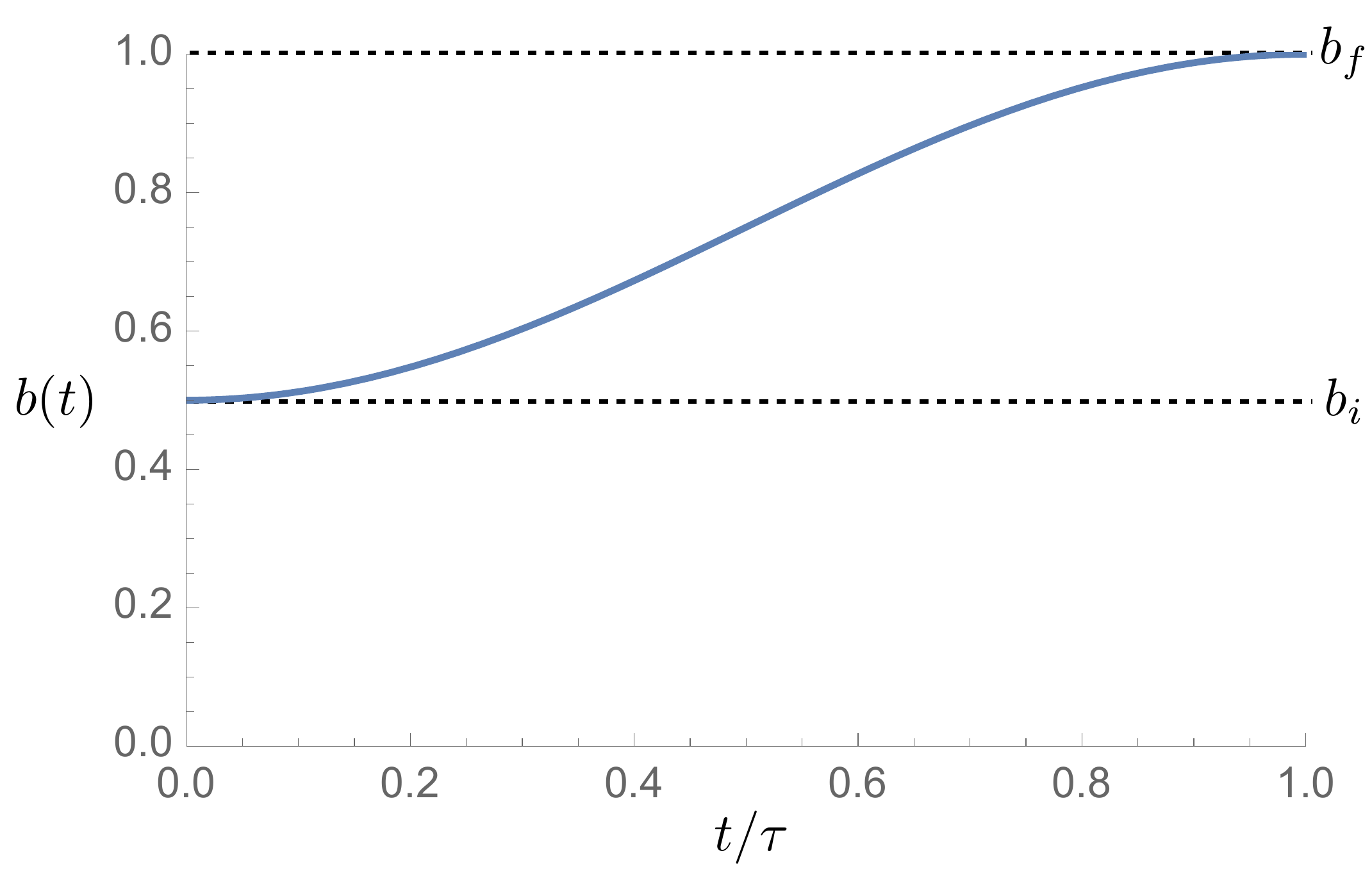}
\caption{Nonlinear protocol for driving between initial bit bias
	$b(0)=b_i$ and final bias $b(\tau)=b_f$. The nonlinear protocol $b(t)$
	(blue curve) has zero slope initially and finally such that the
	counterdiabatic potential vanishes at the protocol's beginning and end.
	}
\label{fig:BiasProtocol}
\end{figure}

Let us consider control protocols for which the bit bias trajectory is given by:
\begin{align}
\label{eq:nlinBitBias}
b(t) = b_i \cos^2(\pi t /2\tau) + b_f \sin^2(\pi t /2\tau)
  ~.
\end{align}
This schedule takes the system monotonically from $b(0)=b_i$ to $b(\tau)=b_f$,
as shown in Fig.~\ref{fig:BiasProtocol}. Since $\partial_t b=0$ at $t=0$ and $t
= \tau$, the counterdiabatic potential vanishes at the initial and final times,
except in the special cases that $b_i$ or $b_f$ are either $0$ or $1$.

\begin{figure}[tbp]
\centering
\includegraphics[width=\columnwidth]{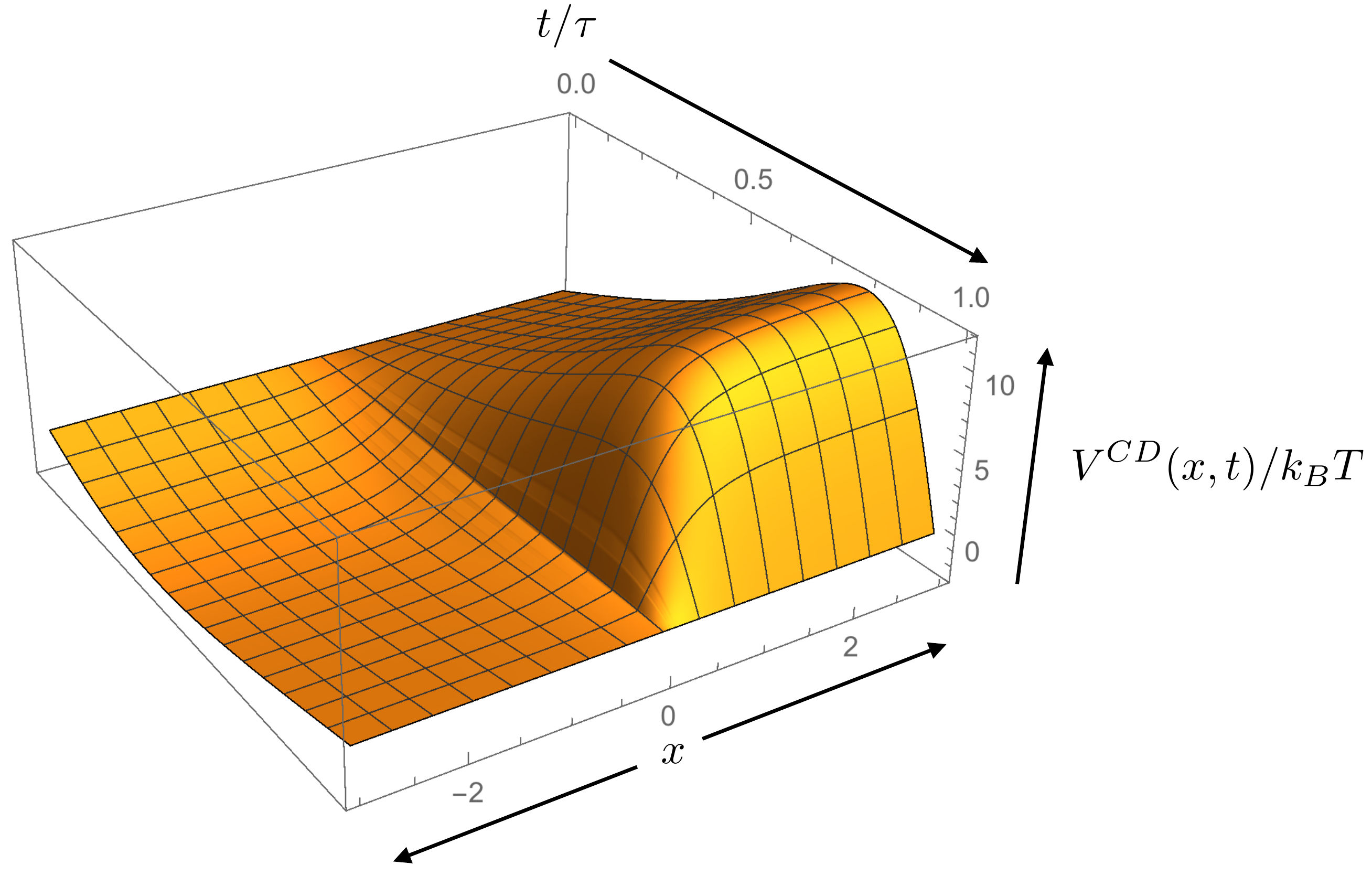}
\caption{Counterdiabatic potential for the nonlinear erasure protocol of Fig.
	\ref{fig:BiasProtocol} which takes a bistable potential well from an
	initial bias $b(0) = 0.5$ to a final bias $b(\tau) = 0.95$. For this
	protocol we set $\tau=1$, $\mu=1$, $k_B T=1$, $\alpha=1$, and $\beta=2$. The
	counterdiabatic potential vanishes at the beginning and end, so that the
	system begins and ends in equilibrium.
	}
\label{fig:VCD}
\end{figure}

The counterdiabatic potential in this case is:
\begin{align}
\label{eq:VCDnonlinear}
V^{CD}(x,t) & = \frac{h(x)}{2 \tau} \\
 & \times \begin{cases}
-\frac{(b_f-b_i)\pi \sin(\pi t/\tau)}{b_i\cos(\pi t/2\tau )^2+b_f\sin(\pi t /2\tau)^2}
  & \text{ if } x\leq 0 \\
  \frac{(b_f-b_i)\pi \sin(\pi t/\tau)}{1-b_i\cos(\pi t/2\tau)^2-b_f\sin(\pi t /2\tau)^2}
  & \text{ if } x>0
\end{cases} \nonumber
  ~.
\end{align}
Note that the explicit dependence on duration factors out, yielding the
prefactor $\tau^{-1}$, as expected. Calculating $h(x)$ numerically, Fig.
\ref{fig:VCD} plots the counterdiabatic potential $V^{CD}(x,t)$. The nonlinear
protocol begins and ends with zero counterdiabatic potential, hence the
distribution begins and ends in equilibrium. This guarantees that when
instantaneously changing back to the default bistable potential landscape, the
work investment beyond the counterdiabatic work equals the change in
nonequilibrium free energy.

\begin{figure}[tbp]
\centering
\includegraphics[width=\columnwidth]{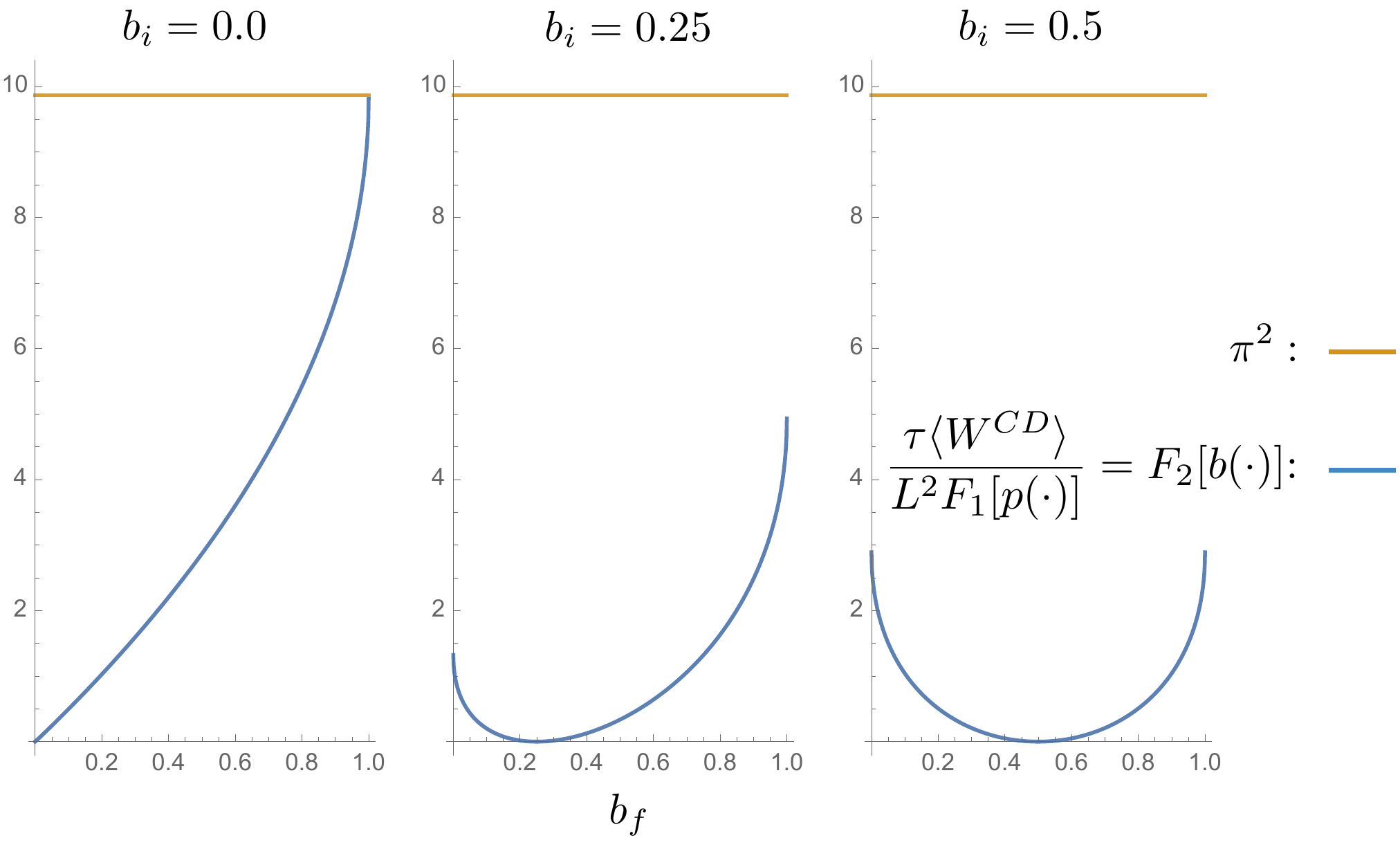}
\caption{Dissipated work to execute a logical operation changes with initial
	bit bias $b_i$ and final bias $b_f$: Dissipated work is proportional to
	$F_2[b(\cdot)]$ when the duration $\tau$ and equilibrium distribution
	$p(\cdot)$ are held fixed. (Left) Initial bit bias $b_i=0.0$: as the
	probability $b_f$ of informational state $0$ increases the cost of erasure
	increases steadily to a maximum at $b_f=1.0$. (Center) Similar behavior for
	an initial bias $b_i=0.25$. (Right) Fair initial distribution $b_i=0.5$,
	demonstrating a finite cost for perfect erasure ($b_f=1.0$) in finite time.
	}
\label{fig:F2}
\end{figure}

Equation (\ref{eq:MetastableVCD}) indicates that any protocol ending with all
probability in a single well, such that $b_f = 0$ or $b_f = 1$, has divergent
counterdiabatic potentials, since either $b(t)$ or $1-b(t)$ vanishes. A
vanishing numerator ${\dot b}(t)$ is no compensation, since under any number of
applications of L'Hopital's rule to evaluate convergence the numerator becomes
nonzero first; it is the derivative of the denominator. Despite this, through
numerical calculations, we find that a counterdiabatic potential $V^{CD}(x,t)$
that starts and ends at zero can perform perfect erasure in finite time with
finite work. If the thermodynamic-computing designer wishes to avoid a
divergent final potential, they can approach perfectly faithful erasure
asymptotically while keeping the final state in equilibrium, because $V^{CD}(x,
\tau)=0$ for all $b_f \neq 0,1$.  As the final bias $b_f$ approaches $1$, the
resulting work approaches a constant value but the system approaches perfect
erasure, as shown in the rightmost plot of Fig. \ref{fig:F2}.

To study the dependence of dissipated work on initial and final bias, $b_i$ and
$b_f$, we substitute the nonlinear bias function, Eq.~(\ref{eq:nlinBitBias}),
into functional $F_2[b(\cdot)]$. This functional is proportional the dissipated
work with the default distribution $p(x)$ and duration fixed. As Fig.
\ref{fig:F2} shows, starting with an unbiased state $b_i= 0.5$, then increasing
the final bias towards $b_f=1.0$, the required work increases, but not
indefinitely. $F^2[b(\cdot)]$ converges to approximately $0.293$, meaning that
the protocol can perform perfect erasure in finite time with finite work.

More generally, Fig. \ref{fig:F2} shows for three different initial biases
$b_i=0.0$, $0.25$, and $0.5$ that the dissipation increases with the magnitude
of the bias difference $|b_i-b_f|$. However, the contribution of the bias
difference is bounded by the case where the bias difference is maximal
$|b_i-b_f|=1$, for which $F_2[b(\cdot)]= \pi^2$.

Note that these plots are intentionally designed in a way similar to Fig. 3 of
Ref. \cite{Zulk14a} and reveal similar dependence on initial and final bias.
Quantitatively, the values are proportional. Reference \cite{Zulk14a} showed
that optimal control in the linear response regime requires dissipated heat
proportional to twice the square of the Hellinger distance:
\begin{align*}
K(b_i,b_f) \equiv \left(\sqrt{b_i}-\sqrt{b_f}\right)^2
  + \left(\sqrt{1-b_i}-\sqrt{1-b_f}\right)^2
  ~.
\end{align*}
Though our chosen bit bias trajectory is not optimal, as discussed in Appendix
\ref{app:Symmetric Metastable Erasure}, numerical integration shows that
the contribution to the dissipated work $F_2[b(\cdot)]\propto K(b_i,b_f)$.
While we do not yet have an explanation of the correspondence between these two
different estimates of dissipated work in finite time, the results' similarity
is suggestive. We should point out, though, that other bias trajectories could
be chosen that are not proportional to the Hellinger distance and may be less
dissipative. It may be a coincidence that our chosen bit bias trajectory
yielded results similar to the linear response analysis of Ref.~\cite{Zulk14a}.

\subsection{Robust Information Storage Requires Work}

With a potential $V(x,0)=\alpha x^4- \beta x^2$ that stores information in
metastable distributional states, there is a finite lifetime of that
information. In this symmetric double well, with one well corresponding to
informational state $0$ and the other to $1$, the lifetime can be quantified in
terms of the average time $\langle \tau_{0 \rightarrow 1} \rangle$ it takes for
a particle to switch between these states. In the overdamped regime this
\emph{information lifetime} is given by~\cite{Kram40a,Hang90a}:
\begin{align}
\langle \tau_{0 \rightarrow 1} \rangle
  = \frac{2 \pi }{\mu \sqrt{| \ddot{V}(x_0,0) \ddot{V}(x_B,0)|}}
  e^{\Delta E_B/\kB T}
  ~,
\label{eq:infoLifetime}
\end{align}
where by $\ddot{V}(x,0)=\partial_x^2V(x,0)$ we denote the curvature of the
default potential energy landscape, $x_0=-\sqrt{\beta/2 \alpha}$ is the
location of the minimum in the metastable $0$ well, $x_B=0$ is the location of
the barrier maximum, and $\Delta E_B=V(x_B,0)-V(x_0,0)$ is the height of the
barrier above the minimum. The latter is a useful measure of the barrier's
energy scale and, thus, how robustly the potential stores information. By
explicit calculation we obtain $\ddot{V}(x_B,0)=-2 \beta$, $\ddot{V}(x_0,0)=4
\beta$, and:
\begin{align}
\Delta E_B = \frac{\beta^2}{4 \alpha}
  ~.
\label{eq:EnergyBarrier}
\end{align}
Hence, the information lifetime is:
\begin{align}
\langle \tau_{0 \rightarrow 1} \rangle
  = \frac{ \pi }{\mu  \beta \sqrt{2}} e^{ \beta^2/4 \alpha \kB T}
  ~.
\end{align}

Note that $\langle \tau_{0 \rightarrow 1} \rangle$ scales as the system length
$L$ squared, due to the $\ddot{V}$ terms in Eq. (\ref{eq:infoLifetime})'s
denominator. Beyond this, the information lifetime is more strongly controlled
by the energy scale of the energy landscape, which can be characterized by
$\Delta E_B$. It has nearly exponential dependence on this energy scale:
\begin{align*}
\frac{\langle \tau_{0 \rightarrow 1} \rangle}{L^2}
  \propto \frac{e^{\Delta E_B/\kB T}}{\Delta E_B}
  ~.
\end{align*}
Thus, we can capture this dependence by evaluating the information lifetime and
scaling by the length. Comparing $f_1[p(\cdot)]$ to $\langle \tau_{0
\rightarrow 1} \rangle$ (i.e.\ $F_1[p(\cdot)]=f_1[p(\cdot)]/L^2$ to $\langle
\tau_{0 \rightarrow 1} \rangle /L^2$) reveals an interesting correspondence
between dissipation and information lifetime, as well as identifying a term
that depends on the default potential's well depth.

\begin{figure}[tbp]
\centering
\includegraphics[width=\columnwidth]{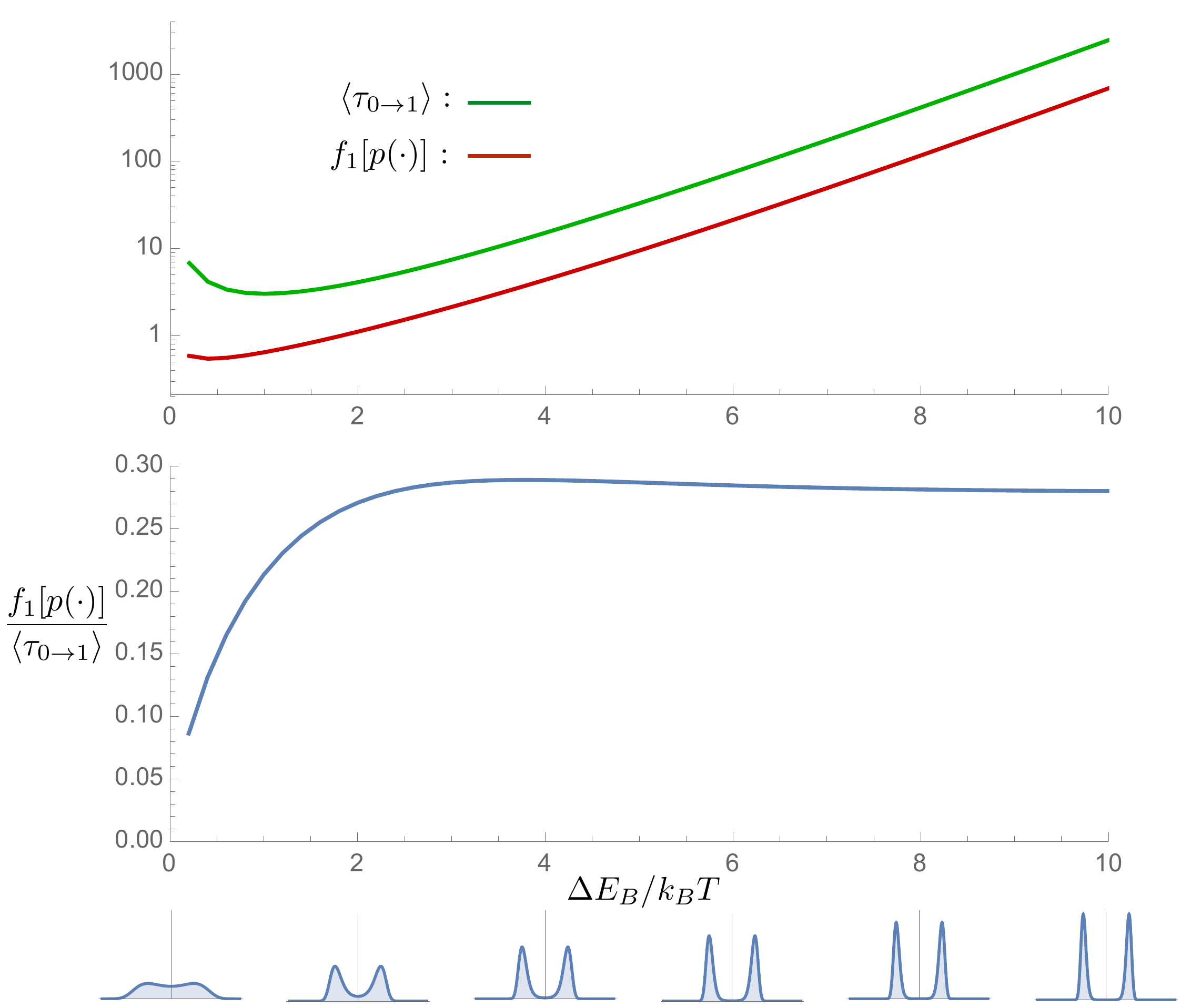}
\caption{Energy barrier dependence: (Top) Changing energy barrier height
	$\Delta E_B$ relative to the thermal energy scale $\kB T$, $f_1[p(\cdot)]$
	and so the required dissipated work increase nearly exponentially. This
	corresponds to an increase in the separation between the distribution in
	the $0$ and $1$ states, as shown by the six distributions along the
	horizontal axis (Bottom). Increasing $\Delta E_B$, greater well separation,
	leads to more robust information storage, as shown by the information
	lifetime $\langle \tau_{0 \rightarrow 1} \rangle$ (Top). (Center) Moreover,
	the information lifetime, which scales just below exponentially, appears to
	scale at the same rate as the dissipated energy when the barrier is at least
	twice $\kB T$. $\kB T =1$, $\alpha =1$, and $\mu =1$ for these
	calculations, while $\beta$ is used to change the energy barrier $\Delta
	E_B$ as in Eq. (\ref{eq:EnergyBarrier}).
	}
\label{fig:F1Comparisons}
\end{figure}

As illustrated in Fig. \ref{fig:F1Comparisons}, with increasing well depth
$\Delta E_B$ the bistable distribution becomes increasingly peaked at the local
minima, and the information lifetime increases nearly exponentially; as
predicted by Eq.~(\ref{eq:infoLifetime}). Interestingly, $f_1[p( \cdot)]$,
which is proportional to the excess work production required during erasure,
scales at roughly the same rate as the information lifetime, if the length
scale is held fixed. As shown in Fig. \ref{fig:F1Comparisons}, the work
required to erase increases nearly exponentially with the height of the energy
barrier between the wells and stays nearly proportional to the information
lifetime. The exception to this occurs for very small barrier heights, where
the potential's equilibrium distribution is not clearly bimodal and there is
nearly unobstructed flow between the information states.

Thus, for finite-time erasure there is a clear energy cost to robust
information storage, which is multiplicatively separable from both the bias
difference, as well as protocol total duration $\tau$. Though we described how
the excess work scales as the square of the length scale $L$, this dependency
is directly contained in the functional $f_1[p(\cdot)]$. This reinforces the
relationship to the information lifetime $\langle \tau_{0 \rightarrow 1}
\rangle$, which also scales as the inverse length scale, due to the term
$\sqrt{|\ddot{V}(x_0,0)\ddot{V}(x_B,0)|}$. However, we also see the direct
effect of the energy-barrier height $\Delta E_B$ through $F_1[p(\cdot)]$ and
its near proportionality to $\langle \tau_{0 \rightarrow 1} \rangle / L^2$.

\section{Conclusion}
\label{sec:Conclusion}

Counterdiabatic control is a new tool for thermodynamic computing that executes
precisely-controlled information processing in finite time at finite cost. It
is implemented via an additional term in the potential energy---the
counterdiabatic potential---which guides the microstate distribution along a
path that results in the desired computation. We analyzed the work required for
counterdiabatic information processing, developing a full suite of resource
trade-offs. Since, as we showed, counterdiabatic protocols are the unique
control that guides the system distribution along a desired trajectory, these
trade-offs apply broadly to any Hamiltonian control in overdamped Fokker-Planck
dynamics in one dimension.  Other than the expected technical complications,
the overall control strategy will generalize to higher-dimensional state
spaces.

We described how to deploy counterdiabatic protocols in combination with
quenching as a general strategy for finite-time metastable information
processing. Since counterdiabatic control exactly specifies the system's final
distribution, it is distinct from previous treatments of finite-time
information processing using geometric control, which focused on driving an
external (thermodynamic) parameter to a desired value with minimal work.

We showed that the work performed during a counterdiabatic protocol separates
into the change in equilibrium free energy $\Delta F^\text{eq}$ and the
counterdiabatic work $\langle W^{CD} \rangle$, which is also the dissipated
work and, thus, proportional to the protocol's entropy production. We showed
that $\langle W^{CD} \rangle$ scales as the inverse of the protocol duration
$\tau$---reinforcing previous analyses of finite-time thermodynamic processes
that showed the work required for information processing increases with
computation rate \cite{Zulk12a, Zulk14a, Lahi16a}. We also showed that
dissipated work scales as the square of the system length scale $L$. That is,
it is more difficult to move distributions long distances in the same finite
time. The time and distance dependence together imply that going twice as far
at the same speed takes twice the energy, rather similar to locomotive
machines.

We then showed that counterdiabatic protocols can process information by adding
quenching at a protocol's beginning and end. Quenching allows rapidly evolving
between nonequilibrium metastable states, which store information. Applying the
approach, we considered a symmetric double-well system and calculated the work
production for various types of finite-time bit manipulation. This analysis
demonstrated that, in addition to the dependence on length scale and duration,
counterdiabatic work depends on erasure fidelity and information storage
robustness.

Evaluating the multiplicative component $F_2[b(\cdot)]$ of the counterdiabatic
work, we found that the dissipation increases with the bit bias difference
between the initial and final distributions. More specifically, it is
proportional to the Hellinger distance for our chosen class of control
protocols. Given an initial equilibrium and unbiased metastable distribution,
the closer the final metastable distribution is to giving all-$0$ informational
states---increased erasure fidelity---the more the operation costs. However,
there is an upper bound on the dissipated work. Thus, it is possible to perform
perfect erasure in finite time at finite cost. It is also possible to flip a
bit in finite time with finite work, as shown in Fig. \ref{fig:F1Comparisons}'s
leftmost plot. Perfect fidelity, though, does not mean results are held
indefinitely.

The factor $f_1[p(\cdot)]$ in the expression for the counterdiabatic work
depends only on the default equilibrium distribution and so it captures the
dependence on information storage robustness. That is, with increased well
depth---and so metastable state robustness---the dissipated work increases
nearly exponentially. Numerical calculations demonstrate that work scales at
the same rate as the information lifetime, which is the Kramers estimate
\cite{Hang90a} of the average time it takes to jump between wells.

A much richer and more detailed picture of resource tradeoffs in thermodynamic
computing emerges. Most concisely, the required work decomposes as follows:
\begin{align*}
\langle W \rangle = \kB T \ln 2 (\H[Y_0]-\H[Y_\tau])
  + \frac{L^2}{\tau}F_1[p(\cdot)]F_2[b(\cdot)]
  ~.
\end{align*}
Landauer's Principle for thermodynamic computing, the first term on the right,
is the work required to reversibly implement a change in metastably-stored
information; it is equal to the change in the physical processor's
nonequilibrium free energy. Counterdiabatic protocols complement and extend
this principle. They reveal, in the second term on the right, an additional
cost in the form of dissipated work that depends on duration $\tau$, length
scale $L$, bias difference through $F[b(\cdot)]$, energy scale of information
storage $\Delta E_B$ through $F_1[p(\cdot)]$, and information lifetime through
the product $L^2 F_1[p(\cdot)]$. The result is a rather more complete picture
of finite-time, accurate thermodynamic computing.

\section*{Acknowledgments}
We thank M. DeWeese, P. S. Krishnaprasad, D. Mandal, P. Riechers, and G.
Wimsatt for helpful discussions.
As an External Faculty member, JPC thanks the Santa Fe Institute and all the
authors thank the Telluride Science Research Center for hospitality during
visits. This material is based upon work supported by, or in part by, John
Templeton Foundation grant 52095, Foundational Questions Institute grant
FQXi-RFP-1609, and the U. S. Army Research Laboratory and the U. S. Army
Research Office under contracts W911NF-13-1-0390 and W911NF-18-1-0028.

\appendix

\section{Uniqueness of Counterdiabatic Protocols}
\label{app:Uniqueness of Counterdiabatic Protocols}

Typically, via the Perron-Frobenius operator, the equations of motion over a
space $\mathcal{X}$ are used to evolve the distribution $\Pr(X_t)$ over states
$x \in \mathcal{X}$ for a time interval $t \in (0,\tau)$ from an initial
distribution $\Pr(X_0)$. The inverse problem, of determining the equations of
motion from the evolution of states, is more challenging. For overdamped
Fokker-Planck dynamics, Ref. \cite{Patr17a} shows how to determine the
counterdiabatic control protocol $V(x,t)=V^Q(x,t)+V^{CD}(x,t)$ directly from
the desired evolution of $\Pr(X^d_t)$ and, hence, determine the equations of
motion. The equations of motion are specified by a changing potential landscape
$V(x,t)$. However, while the counterdiabatic potential is a solution to the
inverse problem, given distribution trajectory $\{\Pr(X^d_t)\}$, such solutions
a priori need not be unique. Here, we show that the counterdiabatic protocol is
the unique protocol that generates the distribution trajectory.

We start by assuming that $V(x,t)$ induces the evolution of $\Pr(X_t)$ over
the time interval $(0,\tau)$. This means that it solves the Fokker-Planck
equation:
\begin{align}
\frac{\partial \Pr(X^d_t=x)}{\partial t}
  &= \mu \frac{\partial}{\partial x}
  \left( \Pr(X^d_t=x)\frac{\partial V(x,t)}{\partial x}\right) \nonumber \\
  & \qquad + \mu \kB T \frac{\partial^2 \Pr(X^d_t=x)}{\partial x^2}
  ~.
\label{eq:Fokker-PlanckSimple}
\end{align}
If the potential is not the unique dynamic solving this equation, then there
exists potential energy landscape:
\begin{align*}
V'(x,t)=V(x,t)+\Delta V(x,t)
  ~,
\end{align*}
that also solves this equation with nonzero $\Delta V(x,t)$. That is:
\begin{align*}
\frac{\partial \Pr(X_t=x)}{\partial t}
  & = \mu \frac{\partial}{\partial x}
  \left( \Pr(X^d_t=x)\frac{\partial V(x,t)+\Delta V(x,t)}{\partial x}\right) \\
  & \qquad + \mu \kB T \frac{\partial^2 \Pr(X^d_t=x)}{\partial x^2}
  ~.
\end{align*}
Subtracting Eq. (\ref{eq:Fokker-PlanckSimple}) gives:
\begin{align*}
0 = \mu \frac{\partial}{\partial x}
  \left( \Pr(X^d_t=x)\frac{\partial \Delta V(x,t)}{\partial x}\right)
  ~.
\end{align*}
Solving for the difference between the two possible solutions leads to the
conclusion that all possible solutions for the difference have the form:
\begin{align*}
\Delta V(x,t)=C(t)+K(t) \int_0^x \frac{dx'}{\Pr(X^d_t=x')}
  ~,
\end{align*}
where $C(t)$ and $K(t)$ can vary with time, but are constant in the positional
variable $x$. $C(t)$ is an expected and trivial additional component:
one can add an additional flat potential to any protocol without physical
consequence beyond the change in total potential between the start and end:
$C(\tau)-C(0)$. However, $K(t)$ corresponds to an additional force, which can
have meaningful affect on the work invested during a control protocol. Thus,
it appears that there are multiple ways to solve for potential energy
underlying the state dynamics. However, if the state space $\mathcal{X}$ is
truly an unbounded spatial degree of freedom, topologically equivalent to the
real numbers $\mathbb{R}$, then the additional solutions corresponding to
nonzero $K(t)$ are unphysical.

Framed another way, these additional components in possible alternative
solutions correspond to addition to the force field $F'(x,t)=F(x,t)+\Delta
F(x,t)$, where:
\begin{align*}
\Delta F(x,t)& =-\frac{\partial \Delta V(x,t)}{\partial x} \\
  & =-\frac{K(t)}{\Pr(X^d_t=x)}
  ~,
\end{align*}
and the force is defined $F(x,t)\equiv-\partial_x V(x,t)$.  While the strength
of this force field varies spatially, its sign is the same for all $x$ at a
given time, meaning that the forces at every point are aligned in the same
direction. This additional force corresponds to an addition to the drift
velocity $v_\text{drift}'(x,t)=v_\text{drift}(x,t)+\Delta v_\text{drift}(x,t)$,
given by:
\begin{align*}
\Delta v_\text{drift}(x,t)= \mu \Delta F(x,t)
  ~,
\end{align*}
where the drift velocity is $v_\text{drift} \equiv \mu F(x,t)$. Finally, this
adds to the probability current $J'(x,t)=J(x,t)+\Delta J(x,t)$.
This turns out to be constant over position:
\begin{align}
\Delta J(x,t) & = \Pr(X_t^d=x) \Delta v_\text{drift}(x,t) \nonumber\\
  & = - \mu K(t)
  ~.
  \label{eq:JK}
\end{align}
This constant probability current cannot be realized in an infinite positional
variable since, despite locally preserving the probability distribution,
probability flows out at one extreme end of the spatial degree of freedom.

To explicitly prove that an additional constant probability current is
impossible in positional space $\mathcal{X}$ topologically conjugate to the
real line $\mathbb{R}$, note that the Fokker-Planck equation Eq.
(\ref{eq:Fokker-PlanckSimple}) is the continuity equation $\partial_t
\Pr(X_t=x) = -\partial_x J(x,t)$.  There is an integral form of this equation,
which relates the change in probability in a region $[x_0,x_1]$ to the
probability current through the boundary of the region:
\begin{align*}
\partial_t \int_{x_0}^{x_1} dx \Pr(X_t^d=x)= J(x_0,t)-J(x_1,t)
  ~.
\end{align*}
In order for $J'(x,t)$ to satisfy the Fokker-Planck equation, it must also
satisfy $\partial_t \int_{x_0}^{x_1} dx \Pr(X_t^d=x)= J'(x_0,t)-J'(x_1,t)$.  So
far, there is no contradiction, since:
\begin{align*}
J'(x_0,t)-J'(x_1,t) & =J(x_0,t)- \mu K(t)-J(x_1,t)+ \mu K(t)
\\ &=J(x_0,t)-J(x_1,t).
\end{align*}
However, in the special case with $x_0=-\infty$---the region of interest is all
$x \leq x_1$---then the only boundary of the region is at $x_1$, such that:
\begin{align*}
\partial_t \int_{-\infty}^{x_1} dx \Pr(X_t^d=x) & =- J(x_1,t)
\\ & = -J'(x,t) \nonumber
  ~.
\end{align*}
For this to be true, $K(t)$ must vanish, and so there cannot be any additional
drift term. That is, up to an additional flat potential $C(t)$, the
counterdiabatic control protocols are the unique way to guide the system along
a desired distribution trajectory $\{\Pr(X_t^d)\}$.

This proof does not preclude additional solutions with nonzero $K(t)$ if the
position variable has circular topology on a finite range $[x_0,x_1]$. This
would mean that $x_0$ and $x_1$ are effectively adjacent such that there can be
probability current at both points. In this case, there are always at least two
boundary surfaces for any region, so it is impossible to use the integral
continuity equation as above. The additional probability current $K(t)$ is
possible, due to probability flow between $x_0$ and $x_1$, which was not
possible between $\infty$ and $-\infty$ in the previous case. However, this
additional probability current corresponds to a force that points in the same
direction along the loop, meaning that system is being driven cyclically. And
so, the dynamics cannot be implemented with Hamiltonian control and must
rely on some free energy resource to be sustained.

\section{Symmetric Metastable Erasure}
\label{app:Symmetric Metastable Erasure}

In metastable erasure, we assume the system is in a metastable distribution of
the initial symmetric equilibrium potential $V(x,0)=V(-x,0)$ during the entire
protocol. If the two metastable informational states are $Y=0$, corresponding
to $x \in (-\infty,0]$, and $Y=1$, corresponding to $x \in (0,\infty)$, then we
can describe a probability distribution trajectory as:
\begin{align*}
\Pr(X_t=x)
  = \begin{cases}
  \Pr(X^\text{eq}=x)2\Pr(Y_t=0) & \text{ if } x\leq 0 \\
  \Pr(X^\text{eq}=x)2\Pr(Y_t=1) & \text{ if } x>0
  \end{cases}
  ~.
\end{align*}
We can then reparametrize in terms of the bit bias, that is
the probability of the $Y=0$ informational state:
\begin{align*}
b(t)=\Pr(Y_{t}=0)
  ~.
\end{align*}
As in Sec.~\ref{sec:Nonequilibrium Information Storage},
let $p(x) = \Pr(X^\text{eq}=x)$ denote the equilibrium distribution,
which inherits the symmetry of the double well potential:
\begin{align*}
p(x) = p(-x)
  ~.
\end{align*}
We then express the evolving metastable
distribution as a function of control parameter:
\begin{align*}
\Pr(X_t=x)
  = \begin{cases}
  2p(x)b(t) & \text{ if } x\leq 0 \\
  2p(x)(1-b(t)) & \text{ if } x>0
\end{cases}
  ~.
\end{align*}

This expression allows us to simplify the
counterdiabatic potential and counterdiabatic work, as follows.
\begin{align*}
& V^{CD}(x,t) =\frac{1}{\mu}\int_{0}^x dx'
  \frac{1}{\Pr(X_t=x')}\int_{-\infty}^{x'} dx'' \partial_t \Pr(X_t=x'') \\
  & =\begin{cases}
	\frac{1}{\mu} \int_{0}^x dx' \frac{1}{2bp(x')}
	\int_{-\infty}^{x'} dx'' 2p(x'') {\dot b} & \text{ if } x\leq0 \\
	\frac{1}{\mu}\int_{0}^x dx' \frac{1}{2(1-b)p(x')}
	(\int_{-\infty}^{0} dx''2 p(x''){\dot b} \\
	\quad + \int_{0}^{x'} dx'' 2p(x'')\partial_t (1-b)) & \text{ if } x>0
  \end{cases} \\
  & =\begin{cases}
	\frac{1}{\mu}\int_{0}^x dx' \frac{1}{bp(x')}
	\int_{-\infty}^{x'} dx'' p(x''){\dot b} & \text{ if } x\leq0 \\
	\frac{1}{\mu}\int_{0}^x dx' \frac{1}{(1-b)p(x')}
	(\int_{-\infty}^{0} dx'' p(x''){\dot b} \\
	\quad -\int_{0}^{x'} dx'' p(-x'') {\dot b}) & \text{ if } x>0
  \end{cases} \\
  & =\begin{cases}
	\frac{1}{\mu}\int_{0}^x dx' \frac{1}{bp(x')}
	\int_{-\infty}^{x'} dx'' p(x''){\dot b} & \text{ if } x\leq0 \\
	\frac{1}{\mu}\int_{0}^x dx' \frac{1}{(1-b)p(x')}
	(\int_{-\infty}^{0} dx'' p(x''){\dot b} \\
	\quad - \int_{-x'}^{0} dx'' p(x'') {\dot b}) & \text{ if } x>0
\end{cases} \\
  & =\begin{cases}
	\frac{1}{\mu}\int_{0}^x dx' \frac{1}{bp(x')}
	\int_{-\infty}^{x'} dx'' p(x''){\dot b} & \text{ if } x\leq0 \\
	\frac{1}{\mu}\int_{0}^x dx' \frac{1}{(1-b)p(x')}
	\int_{-\infty}^{-x'} dx'' p(x''){\dot b} & \text{ if } x>0
\end{cases} \\
  & =\begin{cases}
	\frac{1}{\mu}\int_{0}^x dx' \frac{1}{bp(x')}
	\int_{-\infty}^{-|x'|} dx'' p(x''){\dot b} & \text{ if } x\leq0 \\
	\frac{1}{\mu}\int_{0}^x dx' \frac{1}{(1-b)p(x')}
	\int_{-\infty}^{-|x'|} dx'' p(x''){\dot b} & \text{ if } x>0
\end{cases} \\
  & =\begin{cases}
	\frac{{\dot b}}{b}\frac{1}{\mu}
	\int_{0}^x dx' \frac{1}{p(x')} \int_{-\infty}^{-|x'|} dx'' p(x'')
	& \text{ if } x \leq 0 \\
	\frac{{\dot b}}{1-b}\frac{1}{\mu}
	\int_{0}^x dx' \frac{1}{p(x')} \int_{-\infty}^{-|x'|} dx'' p(x'')
	& \text{ if } x > 0
\end{cases}
  ~,
\end{align*}
where $b=b(t)$ and $\dot b = \partial_t b(t)$.
The second line follows from the first, since $p(x'')=p(-x'')$.

We can substitute $u=-x'$ again since $\int_{0}^x
\frac{1}{p(x')}dx'=-\int_{0}^{-x} du \frac{1}{p(u)}$. And so, if we define:
\begin{align*}
h(x)= \frac{1}{\mu}\int_0^{|x|}dx' \frac{1}{p(x')}
	\int_{-\infty}^{-|x'|}dx''p(x'')
	~,
\end{align*}
then:
\begin{align}
V^{CD}(x,t)
  & = h(x) \times \begin{cases}
	-\frac{{\dot b}}{b}  & \text{ if } x\leq 0 \\
	\frac{{\dot b}}{1-b} & \text{ if } x>0
\end{cases}
  ~.
\end{align}

The resulting counterdiabatic work is:
\begin{align*}
& \langle W^{CD} \rangle = \int_{0}^{\tau} \!\! dt
  \int_{-\infty}^{\infty} \!\! dx \Pr(X_t=x) \partial_t U(x,t) \\
  & = \int_{0}^{\tau} \!\! dt \int_{-\infty}^{0} \!\! dx \,p(x)b\,h(x)
  \partial_t \left(\frac{-{\dot b}}{b} \right) \\
  & \quad + \int_{0}^{\tau} \!\! dt  \int_{0}^{\infty} \!\! dx \,p(x)(1-b)h(x)
  \partial_t \left(\frac{{\dot b}}{1-b} \right) \\
  & = \int_0^\infty  \!\! dx \,p(x) h(x) \times \\
  & \quad \int_0^\tau  \!\! dt \left(b
  \partial_t \left(\frac{-{\dot b}}{b} \right)+(1-b)
  \partial_t \left(\frac{{\dot b}}{1-b} \right)\right) \\
  & = \int_0^\infty \!\! dx \,p(x) h(x) \times \\
  & \quad \int_0^\tau \!\! dt \left(-b
  \left(\frac{\partial_t^2 b}{b}
  -\frac{{\dot b}^2}{b^2}\right)+(1-b)
  \left(\frac{\partial_t^2 b}{1-b}
  +\frac{({\dot b})^2}{(1-b)^2}  \right)\right) \\
  & = \int_0^\infty \!\! dx \,p(x) h(x) \int_0^\tau \!\! dt
  \left(-\partial_t^2 b
  +\frac{({\dot b})^2}{b}+\partial_t^2 b+\frac{({\dot b})^2}{(1-b)}  \right) \\
  & = \int_0^\infty \!\! dx \,p(x) h(x) \int_0^\tau \!\! dt
  \left(\frac{({\dot b})^2}{b}+\frac{({\dot b})^2}{(1-b)}  \right) \\
  & = \int_0^\infty \!\! dx \,p(x) h(x) \int_0^\tau \!\! dt
  \frac{({\dot b})^2}{b-b^2} \\
  & = f_1[p(\cdot)]\times f_2[b(\cdot)]
  ~.
\end{align*}
The second line follows from the first since $p(x)$ and $h(x)$ are symmetric.
The functions appearing on the last line are given by:
\begin{align*}
f_2[b(\cdot)]=\int_0^\tau dt  \frac{({\dot b})^2}{b-b^2}
\end{align*}
and:
\begin{align*}
f_1[p(\cdot)] & =\int_0^\infty dx \,p(x) h(x) \\
  & =\int_0^\infty dx \,p(x) \frac{1}{\mu}\int_0^{|x|}dx'
  \frac{1}{p(x')} \int_{-\infty}^{-|x'|}dx''p(x'') \\
  & =\frac{1}{\mu} \int_0^\infty dx \,p(x) \int_0^{x}dx'
  \frac{1}{p(x')} \int_{-\infty}^{-x'}dx''p(x'')
  ~.
\end{align*}
Thus, the counterdiabatic work is the product of two factors: one dependent on
the bias trajectory $b(t)$, containing all dependence on erasure fidelity, and the other dependent on the equilibrium potential $p(x)$,
containing all dependence on information storage robustness---the information lifetime.

One is tempted to use the functional $f_2$ to find the bias trajectory that
minimizes dissipation. $f_2$ can be expressed as the integral of a Lagrangian:
\begin{align*}
f_2[b(\cdot)]=\int_{0}^{\tau}\mathcal{L}(b(t),b'(t)) dt
  ~,
\end{align*}
where:
\begin{align*}
\mathcal{L}(b(t),b'(t))=\frac{b'(t)^2}{b-b^2}
  ~.
\end{align*}
This implies that, for the optimal path satisfying the equation of motion:
\begin{align}
\frac{\partial \mathcal{L}}{\partial b}=\frac{d}{d t}
  \frac{\partial \mathcal{L}}{\partial b'}
  ~.
\label{eq:optimality}
\end{align}
Integrating these equations of motion, given the constraint of starting
at initial bias $b(0)=b_i$ and ending at final bias $b(\tau)=b_f$ would
determine the most thermodynamically efficient path $b(t)$ for transiting
between different metastable distributions. However, this is challenging
and remains unsolved. So, instead, consider a simpler protocol.

We choose a bias trajectory:
\begin{align*}
b(t)=b(0)\cos(t \pi/2\tau)^2+b(\tau) \sin(t \pi/2\tau)^2
  ~,
\end{align*}
which has vanishing derivative at the protocol's beginning and end, such that
the desired distribution $\Pr(X^d_t)$ has zero time derivative at the initial
and final times. This means that the counterdiabatic potential energy is zero
at the protocol's beginning and end. As a result, the system is in equilibrium at
the beginning and end of the counterdiabatic step in the protocol. Substituting
this into the expression for $f_2[b(\cdot)]$ above, we evaluate numerically,
and see that the counterdiabatic work is proportional to twice Hellinger
distance:
\begin{align*}
K(b_i,b_f) =
  \left(\sqrt{b_i}-\sqrt{b_f}\right)^2+\left(\sqrt{1-b_i}-\sqrt{1-b_f}\right)^2
  ~.
\end{align*}
Despite not knowing the proportionality, we see that the maximum of
$f_2[b(\cdot)]$ occurs when a bit is perfectly flipped, yielding a
multiplicative contribution $\pi^2/\tau$. Substituting our chosen bias trajectory
into the expression
for optimality in Eq. (\ref{eq:optimality}), we see that it does not satisfy
the equality, and so is not optimal.

\bibliography{chaos}

\end{document}